\documentclass[10pt, journal, a4paper, twocolumn]{IEEEtran}
\usepackage[utf8]{inputenc}
\usepackage[arrow, matrix, curve]{xy}
\usepackage{caption}
\usepackage{subcaption}
\usepackage{amsmath}
\usepackage{amsfonts}
\usepackage{amssymb}
\usepackage{float}
\usepackage{hyperref}
\usepackage{tikz}
\usetikzlibrary{decorations.pathreplacing}
\usepackage{graphicx}
\usepackage{float}

\newcommand{\area}{\mbox{area}}
\newcommand{\isom}{\mbox{Isom}}
\newcommand{\eucl}{\mathbb{E}^2}
\newcommand{\hyp}{\mathbb{H}^2}
\newcommand{\sph}{\mathbb{S}^2}
\newcommand{\schl}[2]{$\lbrace #1, #2 \rbrace$}
\newcommand{\im}{\mbox{im }}

\newcommand{\csys}{\mbox{csys}}

\begin{document}

\title{Constructions and Noise Threshold of Hyperbolic Surface Codes}

\author{Nikolas~P.~Breuckmann~%
        and~Barbara~M.~Terhal%
\thanks{Nikolas~P.~Breuckmann. JARA IQI, RWTH Aachen University, 52056 Aachen,
Germany. e-mail: breuckmann@physik.rwth-aachen.de.}%
\thanks{Barbara~M.~Terhal. JARA IQI, RWTH Aachen University, 52056 Aachen,
Germany. e-mail: terhal@physik.rwth-aachen.de.}%
\thanks{Copyright (c) 2014 IEEE. Personal use of this material is permitted.  However, permission to use this material for any other purposes must be obtained from the IEEE by sending a request to pubs-permissions@ieee.org.}
}

\maketitle

\begin{abstract}
We show how to obtain concrete constructions of homological quantum codes based on tilings of 2D surfaces with constant negative curvature (hyperbolic surfaces). This construction results in two-dimensional quantum codes whose tradeoff of encoding rate versus protection is more favorable than for the surface code. These surface codes would require variable length connections between qubits, as determined by the hyperbolic geometry. We provide numerical estimates of the value of the noise threshold and logical error probability of these codes against independent $X$ or $Z$ noise, assuming noise-free error correction.
 \end{abstract}



\section{Introduction} %
\label{sec:introduction}

An essential component of scalable quantum computing is the ability to  protect quantum information reliably from decoherence. Such protection can in principle be achieved via the use of quantum error correcting codes, assuming a sufficiently low physical noise rate and a sufficiently large encoding overhead \cite{quantumcodes_review}. A particularly interesting coding architecture is that of the surface code or the \emph{toric code} first formulated by Kitaev \cite{kitaev:survey, surface_code, freedman_projective, toric_code}. The toric code is an example of a quantum low-density parity check (LDPC) code, meaning that the parity check operators used in quantum error correction only involve a constant number of qubits and each qubit participates only in a constant number of parity check measurements.

The ideas introduced by Kitaev have been generalized to a construction in which a tiled, closed, $n$-dimensional manifold $M$ is translated into a quantum LDPC stabilizer code such that all parity checks are local on $M$. Codes of this type are called \emph{homological stabilizer codes} (see Section \ref{sec:homological_codes}). 
The surface code, based on tiling a euclidean surface, is an attractive code to realize experimentally due to its high noise threshold of almost $1\%$ and the planar qubit layout and inter-qubit coupling (see e.g. recent work in superconducting qubits aimed at making elements of a surface code architecture \cite{Kelly2015,Crcoles2015, Rist2015}). 

\begin{figure}[ht]
  \begin{center}
    \includegraphics[width=0.9\columnwidth]{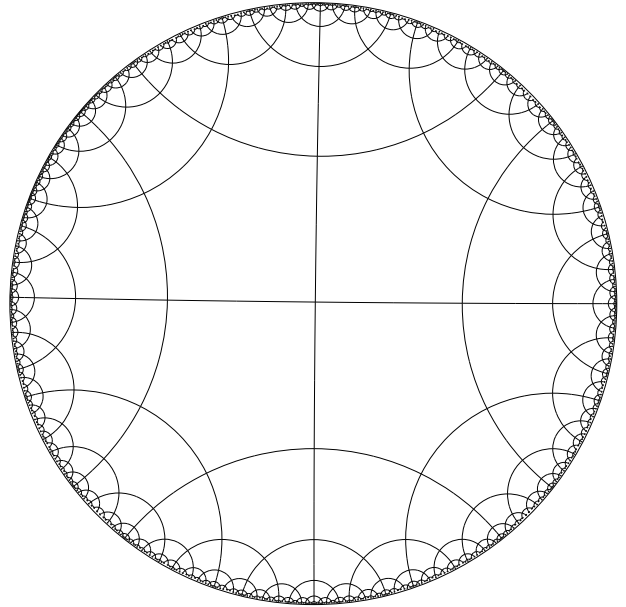} %
  \end{center}
  \caption{The \schl{5}{4}-tiled hyperbolic plane $\hyp$ in the Poincar\'e disc model. Each tile is a pentagon and four tiles meet at a vertex. All lines meet at an angle of $\pi/2$.}\label{fig:poincare_disc}
\end{figure}

However, the surface code also has several non-optimal features which directly result from the fact that it is based on tiling a 2D euclidean manifold. In \cite{no_go_thm} it was shown that any 2D stabilizer code with $n$ physical qubits has distance $d \leq O(\sqrt{n})$ and in \cite{BPT_tradeoffs} it was shown that for any $[[n,k,d]]$ 2D stabilizer code the number of encoded (logical) qubits $k$ and the number of physical qubits $n$ and the distance $d$ obey the trade-off relation $k d^2 \leq c n$ for some constant $c$. Both these results assume that physical qubits are laid out according to an euclidean geometry.
Note that these sets of results do not only pertain to homological stabilizer codes but hold for any LDPC code with a planar qubit lay-out with $O(1)$ distance between qubits.

This rate-distance trade-off bound is achieved by the surface code in which one can, for example, encode $k$ qubits into $k$ separate surface code sheets, each with $d^2$ qubits and distance $d$, leading to a total number of $n=k d^2$ qubits. Clearly, the encoding rate $k/n$ approaches zero when one tries to encode better qubits with growing distance $d$.  

In \cite{fetaya_master_thesis} Fetaya showed that two-dimensional homological codes, based on tilings of arbitrary two-dimensional surfaces, obey the square-root bound on the distance, i.e. $d \leq O(\sqrt{n})$. In \cite{freedman2002z2} it was shown that with a family of homological codes based on 4-dimensional manifolds with non-zero curvature, one can go beyond this square-root bound and encode a qubit with distance scaling as $\sim (n \log n)^{1/2}$. In addition, \cite{freedman2002z2} showed that there exist hyperbolic surface codes which have a constant rate $k/n \geq c_1$ while each encoded qubit has distance scaling as $c_2 \log n$ for some constants $c_1$ and $c_2$. Delfosse \cite{delfosse_tradeoffs} has shown that the logarithmic scaling of the distance is the best one can get for homological codes based on a closed (two-dimensional) surfaces. For such codes he proved that
\begin{align}\label{eqn:tradeoff}
  k d^{2} \leq c (\log k)^{2} n ,
\end{align}
with a constant $c$.
Hyperbolic surface codes could thus offer the possibility of encoding more logical qubits for a given number of physical qubits and a desired level of protection or distance.

However, these codes arise from tilings of a hyperbolic surface, endowed with a hyperbolic distance function, leading to a two-dimensional lay-out of qubits in which qubits interacting in parity check measurements are O(1) distance away with respect to this hyperbolic distance function. If one embeds such a tiling of the surface in an ordinary two-dimensional plane this embedding cannot preserve the distance function (i.e. cannot be isometric). This means that some qubits interacting in parity check measurements will be long-distance as compared to interactions between other qubits.
This non-isometric embedding is demonstrated in the Poincar\'e disk model of the hyperbolic plane shown for a specific tiling in Fig.~\ref{fig:poincare_disc}. One way to visualize the hyperbolic plane $\hyp$ (see more details in Section \ref{sub:curvature}) is by mapping all of its points into the interior of a unit disc (radius $R=1$) within the euclidean plane. The shortest line through two given points, a geodesic, is mapped to an arc of a circle on which both points lie and which intersects the boundary of the disc orthogonally. This mapping preserves angles (it is conformal) but it heavily distorts areas and distances. For example, all pentagons in Fig.~\ref{fig:poincare_disc} have the same hyperbolic area while clearly in the Poincar\'e disk model the distances become exponentially smaller around the boundary (Other models of the hyperbolic plane exist such as the Poincar\'e half-space model in which one represents $\hyp=\{x,y \in \mathbb{R} \mid y >0\}$, i.e. a half-plane, with a metric $ds^2=(dx^2+dy^2)/y^2$).

In the quantum code constructions below, qubits are placed at the edges of the tiles. The tiles themselves represent parity $Z$-checks while tiles on the dual lattice (star operators centered at a vertex) represent $X$-parity checks of the code. In order to encode (multiple) qubits, the hyperbolic surface $\hyp$ has to be closed, similar as a torus is obtained  by closing the euclidean plane by taking periodic boundaries. For the hyperbolic plane $\hyp$, if we want this closing procedure to be isometric, so that locally the surface looks like $\hyp$, it implies that one obtains a torus with many handles (or genus $g > 1$), with each additional handle encoding $2$ additional qubits (see the Gau\ss-Bonnet theorem in Section \ref{sub:properties_of_hyperbolic_surfaces}).
Of course, one can also make a many-handled torus out of a euclidean plane, but this would require some stretching or crumpling; this is another way of saying that the flat curvature of the plane cannot be maintained in the many-handled torus. 

For practical applications it is more interesting to have a fully planar 2D qubit layout such as in the surface code than a code which is defined on qubits distributed over the two-dimensional surface of a many-handled torus. For this reason we consider how to planarize the hyperbolic surface codes in Section \ref{sub:planar_hyperbolic_codes}: we will show that the creation of explicit boundaries on which logical operators have to start and end reduces the efficiency of these codes so that a planar code with constant rate $k/n\geq c_1$ necessarily has a constant distance, no longer increasing (logarithmically) with $n$.

In order to construct a hyperbolic surface code one can pursue the following strategy. First, one constructs a tiling of the hyperbolic plane: such a tiling should be isometric in that it respects this distance measure of the hyperbolic plane, see Section \ref{sec:tilings_of_closed_surfaces}. Such tiling can be associated with a group such that group elements label elementary triangular tiles in which the basic tiles can be decomposed. In order to compactify the hyperbolic plane, i.e. make it into a multi-handled torus which encodes some qubits, one identifies elementary triangular tiles which differ by certain translations: this can be done by finding a normal subgroup of the tiling group. Finding all possible normal subgroups (which normal subgroup one picks determines how many qubits $k$ are encoded) is not simple and has to rely on previous mathematical constructions. In \cite{zemor2009cayley, delfosse_tradeoffs} regular hyperbolic tilings were considered and a construction of finding certain normal subgroups due to Siran was used. In a regular tiling \schl{r}{s} each tile is a regular $r$-gon and $s$ regular $r$-gons meet at each vertex in the tiling (see Section \ref{sec:tilings_of_closed_surfaces}). Ref.~\cite{isaac} considered the $\{7,3\}$ hyperbolic tiling and studied numerically the distance of the logical operators for different encoding rates $k/n$.

Our work expands on these previous constructions by (1) using mathematical tools that allow one to find any compactification of any \schl{r}{s} hyperbolic tiling, (2) analyzing the noise threshold of several interesting tilings and (3) considering the logical error probability and distance of encoded qubits of interesting tilings. Asymptotic scaling of $[[n,k,d]]$ codes is interesting, but in a physical implementation one is interested in using a code with some fixed parameters. We show how a hyperbolic surface code can outperform the regular surface code in terms of encoding more qubits with the same number of physical qubits for a fixed distance (see Table \ref{table:distances} in Section \ref{sub:numerical_analysis}).

 For superconducting transmon qubits, the interaction distance between transmon qubits is flexible when the interqubit coupling is mediated via a transmission line (as in the architecture described in \cite{divincenzo_arch} or the experiments in \cite{Crcoles2015, Rist2015}). The interaction length-scale can vary between $100\: \mu$m (for a compact bus-resonator) to $10$ cm for a transmission line, hence a factor $10^3$.  It may be of interest to consider how to convert the graph generated by the hyperbolic tiling into a planar grid-graph or a double-layer grid-graph with equidistant qubits and minimal non-planar, crossing edges.

In a recent paper \cite{pastawski2015holographic} the authors introduce holographic quantum codes on which physical qubits live at the endpoints of a hyperbolically-tiled surface (for the holographic pentagon code defined in that paper, this is the $\{5,4\}$ tiling) such that these degrees of freedom represent the encoding of logical degrees of freedom which are explicitly entered through the bulk. These codes are thus not directly related to the hyperbolic surface codes analyzed in this paper in which the physical qubits occupy the entire hyperbolic surface. \\

We will review the definition and decoding of homological codes in Section \ref{sec:homological_codes}. In Section \ref{sec:tilings_of_closed_surfaces} we discuss regular tilings and the construction of regular tiled, closed surfaces. We first consider the euclidean case and then generalize to manifolds which admit constant curvature. In Section \ref{sec:constructions} we discuss concrete examples of quantum codes based on closed orientable hyperbolic surfaces with  regular tilings. In Section \ref{sub:numerical_analysis} we give a numerical analysis of the thresholds of families of hyperbolic surface codes. Finally, we discuss the construction of hyperbolic surface codes which are planar (they are embeddable on a disc).

\section{Homological codes} %
\label{sec:homological_codes}
In this section we review how a tiled, closed surface can be used to define a homological (CSS) quantum code using $\mathbb{Z}_{2}$-homology.

\subsection{$\mathbb{Z}_{2}$-homology} %
\label{sub:z_2_homology}

In $\mathbb{Z}_{2}$-homology, statements about the topology of a surface are turned into statements of linear algebra over the field $\mathbb{Z}_{2}=\lbrace 0,1\rbrace$ in which all operations are carried out modulo 2. 

Consider a tiled, closed surface $X$. We assume that two faces overlap on at most a single edge and two edges overlap on at most a single vertex. From now on we will call faces, edges and vertices \emph{2-cells}, \emph{1-cells} and \emph{0-cells}. The tiled surface $X$ is called a \emph{cell complex}.

For a given cell complex $X$ the subsets of $i$-cells in $X$ form a $\mathbb{Z}_{2}$-vector space where the addition of two subsets is given by their symmetric difference (the set of elements which are contained in one set or the other but not in both). The standard basis of this space is given by all sets containing a single $i$-cell. We will identify these sets with the standard basis of $\mathbb{Z}_{2}^{m_{i}}$ where $m_{i}$ is the number of $i$-cells contained in $X$. We denote these vector spaces as $C_{i}(X)$ or simply $C_{i}$ and call their elements \textit{$i$-chains}. For $i=1,2$ one can define a boundary operator $\partial_i$
\begin{align}
	\partial_{i}:C_{i}\rightarrow C_{i-1} 
\end{align}
by its action on the basis vectors: let $e^{i}\in C_{i}$ be a single $i$-cell, then $
\partial_{i}(e^{i})\in C_{i-1}$ is the sum of all $(i-1)$-cells incident to $e^{i}$.

An important property of the boundary operator is that a boundary does not have a boundary 
\begin{align}\label{eqn:boundaryOpHom} 
	\partial_{1}\circ \partial_{2} = 0,
\end{align}
which is equivalent to the statement that boundaries of faces are closed loops. One can also define the dual map called the \textit{coboundary operator} 
\begin{align}
	\delta_{i}: C_{i}\rightarrow C_{i+1} 
\end{align}
which assigns to each $i$-cell all $(i+1)$-cells which are incident to it.\footnote{In this paper we identify chains with cochains via the inner product.} Viewed as matrices, the coboundary operator is the transpose of the boundary operator of one dimension higher 
\begin{align}\label{eqn:coboundaryIsDual}
  \delta_{i} = \partial_{i+1}^{T}. 
\end{align}
One uses both mappings to define subspaces of $C_{1}$. From the boundary operator one obtains the \textit{cycle space} $Z_{1}= \ker \partial_{1}$ and the \textit{boundary space} $B_{1}= \im 
\partial_{2}$. Due to Eqn.~\ref{eqn:boundaryOpHom}, we always have the boundary spaces as subspaces of the cycle spaces $B_{1}\leq Z_{1}$.

Intuitively, the cycle space $Z_{1}$ represents all $1$-chains which have no boundary, i.e. collections of closed loops. Certainly, all boundaries of $2$-chains have no boundary (which is the content of Eqn.~\ref{eqn:boundaryOpHom}) but there might be loops which ``go around'' the surface in a homologically non-trivial way.
We can count the latter ones by introducing the  \textit{first homology group} $H_{1}=Z_{1}/B_{1}$, such that $\dim H_{1}$ is the number of loops that are non-trivial. For an orientable surface, $\dim H_{1}$ is twice the number of its ``handles'' (genus) $g$ of the surface.

The coboundary operator defines the \textit{cocycle space} $Z^{1}=\ker \delta_{1}$ and the \textit{coboundary space} $B^{1}=\im \delta_{0}$ with $B^{1}\leq Z^{1}$. From Eqn.~\ref{eqn:coboundaryIsDual} it follows that 
\begin{align}
	\langle \delta_{i}a,b \rangle_{C_{i+1}}=\langle a,\partial_{i+1}b \rangle_{C_{i}} 
\end{align}
where $\langle \cdot,\cdot \rangle_{C_{i}}$ denotes the standard inner product in $C_{i}$.
Together with Eqn.~\ref{eqn:boundaryOpHom} it follows that the cocycle space is the orthogonal complement of the boundary space and the coboundary space is the orthogonal complement of the cycle space, i.e.
\begin{align}\label{eqn:perpSpaces}
  B^{1} = Z_{1}^{\perp} \text{ and } Z^{1} = B_{1}^{\perp}. 
\end{align}
In other words, the coboundaries are exactly those chains which have even overlap with the cycles and the cocycles are exactly those chains which have even overlap with the boundaries.

The $i$-cells of a tiled surface $X$ correspond to $(2-i)$-cells of the dual tiling $X^{*}$. By linear extension, this gives rise to an isomorphism 
\begin{align}
	*:C_{i}(X)\rightarrow C_{2-i}(X^{*}) 
\end{align}
of the $i$-chains of $X$ to the $(2-i)$-chains of $X^{*}$. We first note that going to the dual chains leaves the inner product (even or oddness of their overlap) invariant 
\begin{align}
	\label{eqn:innerProdInvariant} \langle a,b \rangle_{C_{i}} = \langle *a,*b \rangle_{C^{*}_{2-i}}. 
\end{align}
Directly from their definitions, it is also clear that applying the coboundary operator to a chain of a cell complex is equivalent to going to the dual complex and applying the boundary operator of the complementary dimension
\begin{align}\label{eqn:coboundDualToBound}
  \delta_{i} = *^{-1} \circ \partial_{2-i} \circ * 
\end{align}
Or in diagrammatic form: 
\begin{align}
	\begin{split}
		\xymatrix{ C_{i} \ar[r]^{\delta_{i}} \ar[d]_{*} & C_{i+1} \ar[d]^{*} \\
		C^{*}_{2-i} \ar[r]_{
		\partial_{2-i}} & C^{*}_{2-i-1} } 
	\end{split}
\end{align}

\begin{figure}[hb]
  \begin{center}
\includegraphics[scale=1]{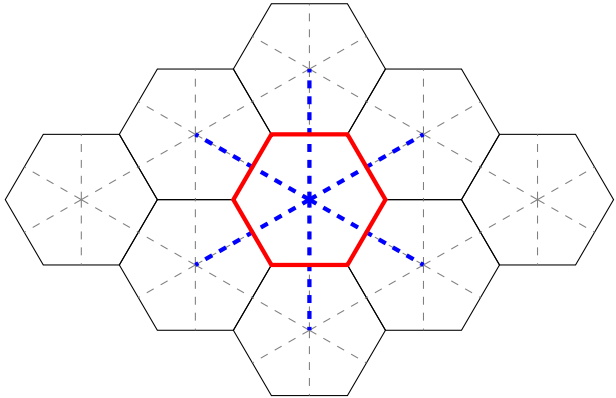} 
  \end{center}
  \caption{Part of the \schl{6}{3}-tiling. The dual tiling (dashed) is the \schl{3}{6}-tiling. The boundary of a face (red) corresponds to the coboundary of a vertex in the dual structure (blue).} \label{fig:dual}
\end{figure}

Eqn.~\ref{eqn:coboundDualToBound} allows us to interpret the coboundaries $B^{1}$ and cocycles $Z^{1}$ as boundaries and cycles of the dual structure (Fig.~\ref{fig:dual}).

\subsection{Quantum codes} %
\label{sub:quantum_codes}

\begin{figure}
	\begin{center}
\includegraphics[scale=1]{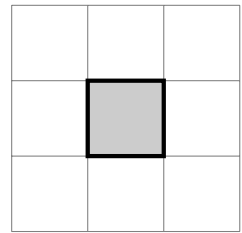} 
	\end{center}
	\caption{A \schl{4}{4}-tiling of the torus (opposite sides are identified). The 1-chain highlighted in black is the boundary of the gray square, but also the boundary of all white squares.} \label{fig:linDep} 
\end{figure}

From the $Z_2$-homology of a tiled surface, we can define a so called \emph{stabilizer code} encoding $k$ logical qubits into $n$ physical qubits, $\mathcal{H} = (\mathbb{C}^2)^{\otimes n}$ (see background on stabilizer codes in \cite{quantumcodes_review}). The $2^k$-dimensional codespace $C \subseteq \mathcal{H}$ is defined as the $+1$ eigenspace of an Abelian subgroup $S$ ($-I \notin S$) of the Pauli group.
To turn a closed, tiled surface $X$ into a stabilizer code we identify all its edges with qubits. The boundaries of the faces are used to define $Z$-type check operators (stabilizer elements). For every face we add a generator to $S$ which acts as $Z$ on all edges which belong to the boundary of the face. Equivalently, the set of coboundaries of vertices gives a generating set of $X$-check operators. Note that in general the generating set of $Z$- and $X$-stabilizers is not independent (Fig.~\ref{fig:linDep}). From now on, when we refer to \textit{the} stabilizer generators we mean this canonical set associated to the faces and vertices.

We call the number of physical qubits on which a Pauli operator acts non-trivially its \emph{weight}. If the surface is tiled by a regular tiling given by the Schl\"afli symbol \schl{r}{s} (see definition in Section \ref{sec:tilings_of_euclidean_surfaces}), then $r$ is the weight of the $Z$- and $s$ the weight of the $X$-stabilizers. 

The number of \textit{physical qubits} is simply the number of edges $n=\dim C_{1}$. The number of \textit{encoded qubits} can be calculated by taking the number of physical qubits and subtracting the number of restrictions that the stabilizers impose:
\begin{align}
	\begin{split}
		k &= \dim C_{1} - \dim B^{1} - \dim B_{1} \\ &= \dim C_{1} - \dim Z_{1}^{\perp} - \dim B_{1} \\
		&= \dim C_{1} - (\dim C_{1} - \dim Z_{1}) - \dim B_{1}\\ &= \dim H_{1} 
	\end{split}
\end{align}

There is another way of seeing this which helps the intuitive understanding of homological codes: Consider the subgroup $N(S)$ of the Pauli group which consists of those operators which commute with all elements in $S$.\footnote{The $N$ stands for normalizer. As Pauli operators either commute or anti-commute $N(S)$ coincides with all elements that commute with $S$.} The action of $N(S)$ on $\mathcal{H}$ leaves the space of logical qubits as a whole invariant. The $Z$- and $X$-type elements of $N(S)$ stand in one-to-one correspondence with the cycles and cocycles by Eqn.~\ref{eqn:perpSpaces}. All stabilizers (the boundaries and coboundaries) have a trivial action on the logical qubits. The operators which are cycles but not boundaries, i.e. non-trivial loops, have a non-trivial action on the qubits. With the help of Eqn.~\ref{eqn:coboundaryIsDual} one can see that cycles which are not boundaries and cocycles which are not coboundaries come in $\dim H_{1}$ pairs with odd overlap, whereas the overlap of elements from different pairs is even. We identify these pairs of non-trivial loops as the $X$- and $Z$-operators, each acting on one of the logical qubits.

The \textit{distance} $d$ of the code is given by the minimum weight of a logical operator. By the previous discussion this is the same as the minimum length of a homologically non-trivial cycle in the tiling or its dual. This quantity is also known as the \textit{combinatorial systole} and denoted by $\csys$ (or $\csys^{*}$ for the dual case).

\begin{figure}[hb]
	\begin{center}
		\begin{tabular}
			{|c|c|} \hline \textbf{Stabilizer code} & \textbf{Homology} \\
			\hline $S_Z$ & $B_1$ \\
			\hline $S_X$ & $B^1$ or $B^{*}_{1}$ \\
			\hline $N(S_X)_Z$ & $Z_1$ \\
			\hline $N(S_Z)_X$ & $Z^1$ or $Z_{1}^{*}$ \\
			\hline $n$ & $\dim C_1$ \\
			\hline $k$ & $\dim H_1$ \\
			\hline $d$ & $\min (\csys,\csys^{*})$ \\
			\hline 
		\end{tabular}
	\end{center}
	\caption{Overview: Corresponding notions in the language of stabilizer codes and homology. The subscript $X$ and $Z$ restricts the set to $X$- respectively $Z$-type operators.} \label{fig:stabHomCorrespondence} 
\end{figure}

\subsection{Error correction in homological codes} %
\label{sub:error_correction_in_homological_codes}

A stabilizer code is used to actively correct errors by measuring the generators of $S$. The measurement result is called the \emph{syndrome}. The syndrome information is then used to infer a recovery operation. Note that the error only needs to be corrected up to a stabilizer element.

\begin{figure}[ht]
	\begin{center}
\includegraphics[scale=1]{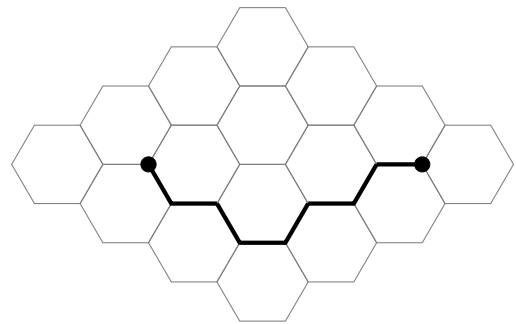} 
	\end{center}
	\caption{The error chain $E_{Z}$ is highlighted. At its boundary we have two vertices which constitute the syndrome of the error.} \label{fig:errorChain} 
\end{figure}

For homological codes, error correction can be understood geometrically. Every  Pauli error can be identified with two chains $E_{X}, E_{Z}\in C_{1}$, where the support of each chain tells us where the error acts as $X$ respectively $Z$ on the qubits. Consider for concreteness $E_{Z}$. The syndrome is the set of all $X$-stabilizers which anti-commute with the error $E_{Z}$ (Fig.~\ref{fig:errorChain}). Since we identify $X$-stabilizers with vertices the syndrome is simply the support of $\partial E_{Z}$. Analogously, the syndrome of the $X$-error $E_{X}$ is given by $\delta E_{X}$ or equivalently by the boundary of $*E_{X}$ in the dual.

A $Z$-error $E_{Z}$ is corrected by applying a $Z$-type Pauli operator with support $R\in C_{1}$ such that its boundary coincides with the syndrome. After this is done the system is back in a code state. If the sum of chains $E_{Z}+R$ is in $B_1$, we applied a stabilizer and no operation was performed on the logical qubits. If, however, $E_{Z}+R$ contains a non-trivial loop, then we applied a logical operator and our encoded information is corrupted. An $X$-error is corrected in the exact same way but in the dual tiling.

In general, a syndrome will consist of many vertices of the (dual) lattice. The problem of correcting the error is choosing and connecting pairs such that the likelihood of a resulting logical operator is minimized. This depends on the probabilities for different errors to occur. Those are in turn determined by the \emph{error model}. Typically, errors with high weight are assigned a low probability. Hence, a good recovery strategy is to find the lowest-weight chain $R$ consistent with the syndrome. This problem is a variant of \emph{minimum weight (perfect) matching} (MWM) which can be solved efficiently by the so-called blossom algorithm.

\section{Tilings of closed surfaces} %
\label{sec:tilings_of_closed_surfaces}

\subsection{Euclidean tilings} %
\label{sec:tilings_of_euclidean_surfaces}

In this section we introduce tilings of the euclidean plane (denoted by $\eucl$). The concepts introduced in this section will directly carry over to spaces with constant curvature.

The set of all distance-preserving mappings from the euclidean plane onto itself forms a group under composition. This group is called the \emph{group of isometries} $\isom(\eucl)$. Its elements are either translations, rotations or combinations of translations and reflections (called \emph{glide reflections}).
Let $G$ be a subgroup of $\isom(\eucl)$. For a point $P\in \eucl$, we call the set of all points that $P$ is mapped to by elements of $G$ the \emph{orbit} of $P$ under $G$.

Take for example the group of all integer translations to the right. The orbit of a point $P=(x,y)\in \eucl$ is the set $\lbrace (x+n,y)\mid n\in \mathbb{Z}\rbrace$.

A \emph{fundamental domain} of a group $G$ is a part of the plane which contains a representative of each $G$-orbit with at most one representative of each $G$-orbit in its interior. For the aforementioned example a choice for a fundamental domain is the strip $[0,1]\times \mathbb{R}$.

The action of $G$ on the points of $\eucl$ induces an action on a fundamental domain $F$ of $G$. We denote the application of a $g\in G$ on $F$ by $F*g$. The identity element $e\in G$ acts as the identity map $F*e=F$ and the action is compatible with the group multiplication. By the latter we mean that the application of $gh$ on $F$ is the same as first applying $g$ to $F$ and then $h$ to the result, or in short $F*(gh)=(F*g)*h$.

If we have a set of (finite) parts of the plane which cover $\eucl$ and all parts either do not overlap or overlap on their boundaries we call this a \emph{tiling} of $\eucl$. Each element of the tiling is called a $tile$ or $face$. The orbit of a fundamental domain naturally gives rise to a tiling.

We will focus on \emph{regular tilings} where all faces are (congruent) regular polygons which are placed edge-to-edge. Each regular tiling can be labeled by the number of sides of the polygons $r$ and the number of polygons $s$ meeting at a point. This label is known as the \emph{Schläfli symbol} \schl{r}{s}. Every \schl{r}{s}-tiling has a well defined dual tiling with Schläfli symbol \schl{s}{r} (Fig.~\ref{fig:dual}).

We will now discuss \emph{Wythoff's kaleidoscopic construction} which allows us to relate a regular tiling \schl{r}{s} to a group of isometries $G=G_{r,s}$.

\begin{figure}
    \centering
    \begin{subfigure}[b]{0.45\textwidth}
        \centering
\includegraphics[scale=1]{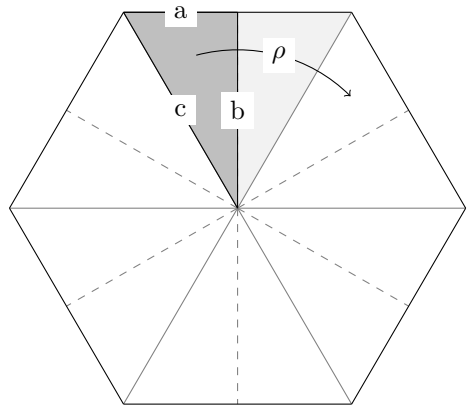} 
        \caption{Action of the symmetry group on a single face.}
        \label{fig:sym_face}
    \end{subfigure}
    \hfill
    \begin{subfigure}[b]{0.45\textwidth}
        \centering
\includegraphics[scale=1]{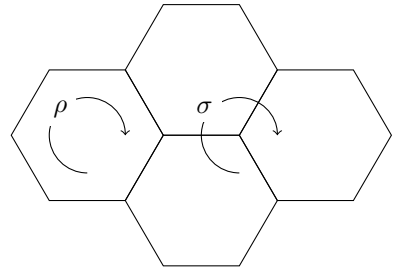} 
        \caption{Rotations acting on the lattice}
        \label{fig:sym_lattice}
    \end{subfigure}
    \caption{Group acting on the \schl{6}{3}-tiling.}
    \label{fig:group_action}
\end{figure}

A regular $r$-gon has $2r$ symmetries generated by the reflections on its symmetry axes (Fig.~\ref{fig:sym_face}). The symmetry axes induce a triangulation of the $r$-gon into $2r$ (right) triangles. The triangulation of the faces induces a triangulation of the whole tiling into triangles with internal angles $\pi/2, \pi/r$ and $\pi/s$.

Let $G_{r,s}$ be generated by the reflections on the edges $a,b,c$ of a single triangle. We assume that $a$, $b$ and $c$ are arranged in clockwise order such that $a$ is opposite to the angle $\pi/r$ and $b$ is opposite to the angle $\pi/s$.

A reflection applied twice is the same as doing nothing, so that $a^2=b^2=c^2=e$. Additionally, two consequtive reflections on two lines that intersect at an angle $\pi/k$ for some positive integer $k$ correspond to a rotation around the intersection point by an angle $2\pi/k$. Thus, $k$ rotations give the identity.

These are all the relations that $a$, $b$ and $c$ fulfill and the multiplication of any two elements of $G_{r,s}$ is completely determined by them. Hence, we can write compactly
\begin{align}
  G_{r,s} = \langle a,b,c \mid a^2=b^2=c^2=(ab)^2=(bc)^r=(ca)^s=e \rangle .
\end{align}
This is called the \emph{presentation} of the group $G_{r,s}$. One way to think about group presentations is as a set of strings or words consisting of $\lbrace a,b,c, a^{-1},b^{-1},c^{-1} \rbrace$. Group multiplication simply corresponds to a concatenation of words. All words which differ by a subword that is equivalent to $e$ by the rules of group multiplication (such as $g^{-1}g=e$) or any of the given relations are considered to be equal. Abstract group presentations can be directly used in computer algebra systems such as \textsc{GAP} or \textsc{Magma}.
 
There is an important subgroup of $G_{r,s}$ which we will denote by $G_{r,s}^{+}$. It consists of all orientation-preserving maps (rigid motions). The group $G_{r,s}^{+}$ is generated by all double reflections $\rho = bc$ and $\sigma = ca$, so that
\begin{align}\label{eqn:gamma_plus}
  G_{r,s}^{+} = \langle \rho, \sigma \mid \rho^{r}=\sigma^{s}=(\rho \sigma)^{2}=e \rangle .
\end{align}
The generators $\rho$ and $\sigma$ act as a clockwise $2\pi/r$ and $2\pi/s$ rotations, respectively. Note that the $G_{r,s}^{+}$-orbit of a triangle only covers ``half'' the plane as we cannot map between triangles that share an edge. We fix this by taking two triangles related by a $b$-reflection as the fundamental domain of $G_{r,s}^{+}$.

The important observation to make is that (by construction) each element of the orbit of the group acting on a fundamental domain is uniquely labeled by a group element. By fixing a triangle $F_{0}$ that is the fundamental domain of $G_{r,s}^{+}$, every other triangle $F$ can be labeled by the group element $g\in G_{r,s}^{+}$ that maps $F_{0}$ onto it, i.e. $F_{0}*g = F$. This allows us to forget about the group action ($F_{0}$ was arbitrary in the first place) and only talk about $G_{r,s}^{+}$.

Let $\langle \rho \rangle = \lbrace e, \rho, \rho^{2},\dotsc,\rho^{r-1}\rbrace$ be the cyclic subgroup of $G_{r,s}^{+}$ that is generated by the rotation $\rho$. The faces of the \schl{r}{s}-tiling (the $r$-gons) are in one-to-one correspondence to the triangles \emph{up to a rotation by $\rho$}. In group theoretic language these are the \emph{left cosets} of the subgroup $\langle \rho \rangle$ denoted by $g\langle \rho \rangle=\lbrace g, g\rho, g\rho^{2},\dotsc,g\rho^{r-1} \rbrace$ for a $g\in G_{r,s}^{+}$. Similarly the vertices and edges of the tiling can be uniquely labeled by left cosets of the cyclic subgroups $\langle \sigma \rangle$ and $\langle \rho\sigma \rangle$, respectively.

Note that faces, edges and vertices are incident if and only if their associated cosets share a common element. This means that the topology of the lattice is  encoded in the group $G_{r,s}^{+}$ as well.

\subsection{Quotient surfaces} %
\label{sub:quotient_surfaces}

In this section we review how to use isometries to define closed, two-dimensional manifolds (or surfaces) which locally look like $\eucl$. 

Let $H$ be a subgroup of isometries acting on $\eucl$. We can construct a new space $\mathbb{E}^{2}/H$ where we identify all points that differ by the application of an element of $H$. Formally, each point is now a set of the form $\lbrace P*g\mid g\in H \rbrace$, i.e. the orbit of some point $P$.

To avoid degenerate cases, one demands that the action of $H$ does not have any fixed points. Additionally, we assume that there exists an $\epsilon > 0$ such that for all points $P$ in $\eucl$ the discs of radius $\epsilon$ around every point in the orbit $P*H$ do not overlap. If $H$ meets these requirements one says that it acts \emph{fixed-point free} and \emph{discontinuous}. In this case there exists a fundamental domain of $H$ around every point of $\eucl$.

In an earlier example we considered a group $H$  consisting of all integer translations along the $x$-axis. $H$ acts fixed-point free and discontinuous and its quotient surface $\eucl/H$ is an infinite cylinder with circumference one. One can imagine that $H$ wraps $\eucl$ up in its $x$-direction and positions all points belonging to one orbit `on top of each other'. In this sense $\eucl$ is a \emph{covering} of the surface $\eucl/H$ and $H$ is called the \emph{covering group}.

Note that in the example the geometry of the cylinder is the same as the geometry of the plane within discs of radius $< 1/2$ . This radius is called the \emph{injectivity radius} $R_{inj}$. The injectivity radius plays an important role in code construction of Section \ref{sec:homological_codes} as it provides a lower-bound on the length of the shortest closed loop on $\eucl/H$ that cannot be contracted to a point.

For a quantum code it follows that if the tiling is defined on a quotient surface then all loops contained within a disc with radius $R_{inj}$ must be boundaries. This is because the geometry within the disc is the same as the geometry of $\eucl$ and all loops in $\eucl$ are boundaries of the set of faces that they surround. The injectivity radius therefore provides a lower bound on the code distance $d$, assuming that all edges are of unit length.

Which quotient surfaces $\eucl/H$ admit a regular tiling? One condition on $H$ is, that it needs to respect the tiling structure. This means that it must be a subgroup of the tiling group $G_{r,s}$.

The faces, edges and vertices of the tiling on the quotient surface $\eucl/H$ are similarly labeled by the action of $G_{r,s}$ on an \emph{arbitrarily} chosen face. However, for this to work on the quotient surface it must ``look the same'' everywhere and in every direction. The action of the covering group $H$ should not affect the labels. This is the case if for all $h\in H$ and $g\in G_{r,s}$ we have $ghg^{-1}\in H$. We say that $H$ is a \emph{normal subgroup} of $G_{r,s}$.

If $H$ does not contain any glide-reflections ($H$ is a subgroup of $G_{r,s}^{+}$), its quotient surface $\eucl/H$ is orientable.

In the tiling of $\eucl/H$, each face can be labeled by a set of the form $g\langle\rho\rangle H =  \lbrace g\rho^{n} h \mid n\in \lbrace 1,\dotsc,r\rbrace, h\in H\rbrace$ for a $g\in G_{r,s}^{+}$. Similarly, we have a labeling of the edges and vertices of the quotient surface $\eucl/ H$ using cosets of $\langle\rho\sigma\rangle H$ and $\langle\sigma\rangle H$. Faces, edges and vertices of $\eucl/H$ are incident if and only if their associated cosets share a common element.

The number of cosets of a subgroup in a parent group is called the \emph{index} of a subgroup. If the index of $H$ in $G_{r,s}$ is finite then $\eucl/H$ has finite area.

In conclusion, one is able to define a group of isometries $G_{r,s}$ which encodes a regular tiling \schl{r}{s}. The problem of finding closed, euclidean, orientable surfaces which admit a \schl{r}{s} tiling reduces to finding finite index, normal subgroups of $G_{r,s}$ which have no fixed-points.

\subsection{Surfaces with Curvature} %
\label{sub:curvature}

The curvature of an arc of radius $R$ can be measured by $1/R$. We can likewise measure the curvature of a surface at a point $P$ by cutting it with two normal planes to create two arcs with radii $R_1$ and $R_2$. For a sphere, the two resulting arcs have their radii of curvature on the same side. For a saddle, the radii of curvature can show in different directions. In this case we assign the radii different signs. We define the curvature $K$ at point $P$ by taking the maximum and minimum possible values of $R_{1}$ and $R_{2}$ and setting $K=1/R_{1}R_{2}$. Hence, the curvature of a sphere with radius $R$ is $K=1/R^{2}$ at any point. For a saddle, $K$ will be negative and for the euclidean plane $K$ vanishes. $K$ is called the \emph{Gaussian curvature}.

Any triangle on the unit sphere $\sph$ has angles which sum up to more than $\pi$ due to the curvature. This is called an \emph{angular defect}. We can define a regular tiling on $\sph$ using Wythoff's construction that we introduced in Section \ref{sec:tilings_of_euclidean_surfaces}. A regular $r$-gon on $\sph$ can be divided into $2r$ triangles. For a regular tiling \schl{r}{s} the internal angles of the triangles are $\pi/2$, $\pi/r$ and $\pi/s$. To make these angles to add up to more than $\pi$ we have the condition that $1/r+1/s > 1/2$. The only regular tilings which fullfill this condition are \schl{3}{3}, \schl{3}{4}, \schl{4}{3}, \schl{3}{5} and \schl{5}{3}. These tilings are well known as they correspond to the Platonic solids.

Similarly, for the euclidean plane we have the condition that for a \schl{r}{s}-tiling it has to hold that $1/r+1/s = 1/2$ as all angles need to sum up to $\pi$. Hence, \schl{6}{3}, \schl{3}{6} and \schl{4}{4} are the only regular tilings of $\eucl$.

If the curvature is negative the internal angles have to sum up to a value smaller than $\pi$. The condition $1/r+1/s < 1/2$ is now met for an infinite number of pairs $r,s$. The plane with constant, negative curvature $K=-1$ is called the \emph{hyperbolic plane} $\hyp$ (see Fig. 7).

\begin{figure}[ht]
  \begin{center}
   \includegraphics[width=0.8\columnwidth]{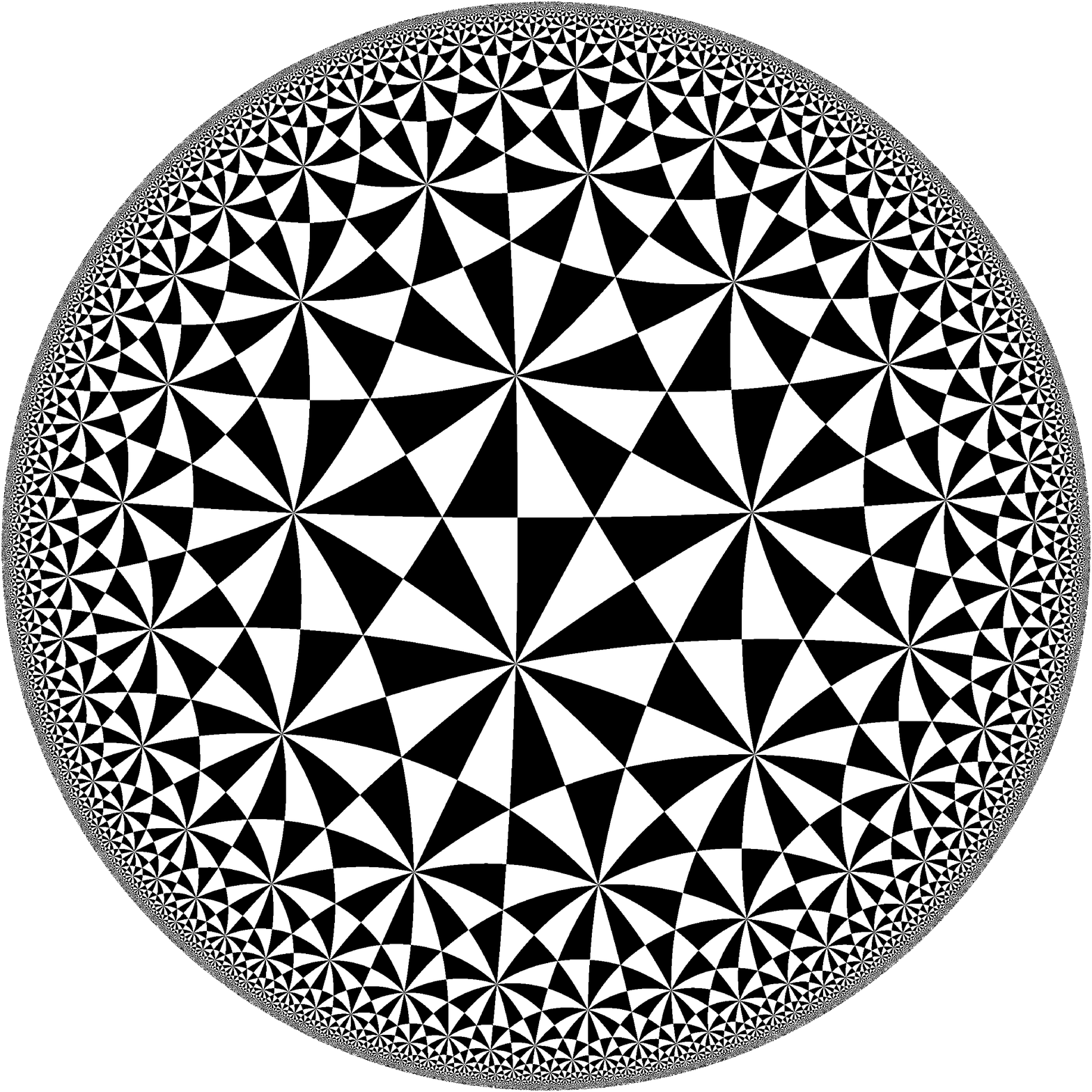} %
  \end{center}
\caption{The \schl{7}{3}-tiled hyperbolic plane $\hyp$ in the Poincar\'e disc model. The fundamental domains of $G_{7,3}$ are filled in black and white. Fundamental domains of the same color are related by an element of $G^{+}_{7,3}$.}\label{fig:poincare_disc2}
\end{figure}

All constructions introduced in Section \ref{sec:tilings_of_euclidean_surfaces} can also be applied to the sphere $\sph$ and the hyperbolic plane $\hyp$. We then obtain (regularly tiled) quotient surfaces with curvature $K=+1$ and $K=-1$, respectively.

Curvature does not only determine what regular tilings are possible but it also determines the topology of the quotient surfaces. The sphere itself only has contractible loops. The only quotient surface it admits is called the \emph{projective plane}, which is non-orientable. It is obtained by factoring out a reflection across a plane which intersects the sphere at a great circle. Homological quantum codes derived from tilings of the projective plane were analyzed in \cite{freedman_projective}.

Since the euclidean plane has no curvature, all translations commute. Therefore two non-colinear translations suffice to make the resulting quotient surface closed. If the covering group $H$ contains translations only, the surface $\eucl/H$ is a torus. If $H$ also contains a glide-reflection, the quotient surface $\eucl/H$ is a Klein bottle. These are the only two closed surfaces with vanishing curvature.

We will see in this section that the hyperbolic plane admits more freedom and allows for quotient spaces which have high genus.

We end this section by mentioning that (up to scaling) all closed surfaces of constant curvature can be expressed as quotient surfaces $\sph/H$, $\eucl/H$ or $\hyp/H$ according to the \emph{Killing-Hopf theorem} \cite{stillwell}.

\subsection{Properties of hyperbolic surfaces codes} %
\label{sub:properties_of_hyperbolic_surfaces}

In this section we will analyze the general properties of closed, hyperbolic surfaces and the quantum codes obtained from them.

The \emph{Gau\ss -Bonnet theorem} connects the curvature of surfaces to their topology. It states that for a closed, oriented surface $M$ the integral over the Gaussian curvature $K$ is proportional to the genus $g$ of $M$
\begin{align}
  \int_{M} K dA = 2\pi (2-2g) .
\end{align}
Hence, negative curvature of $M$, $K=-1$ at any point enforces a proportionality between the genus and the area:
\begin{align}\label{eqn:genus_area}
  2g = \frac{\area(M)}{2\pi} + 2.
\end{align}

It is well-known \cite{ratcliffe} that the area of a regular, hyperbolic $r$-gon $\Pi$ is given by its angular defect
\begin{align}
  \area(\Pi) = (r-2)\pi - \sum_{i=1}^{r} \alpha_{i}
\end{align}
where $\alpha_{i}$ are the internal angles of $\Pi$. If $M$ is tiled by \schl{r}{s} we have $\alpha_{i} = 2\pi/s$ and thus
\begin{align}\label{eqn:area_hyp_polygon}
  \area(M) = \pi F \left( r-2-2 \frac{r}{s} \right)
\end{align}
where $F$ is the number of faces in the tiling of $M$.

By substituting Eqn.~\ref{eqn:area_hyp_polygon} in Eqn.~\ref{eqn:genus_area} we see that a $[[n,k,d]]$ quantum code derived from the \schl{r}{s}-tiling of $M$ will have a rate
\begin{align}\label{eqn:hyp_rate}
  \frac{k}{n} = 1 - \frac{2}{r} -\frac{2}{s} + \frac{2}{n}.
\end{align}

The asymptotic rate ($n\rightarrow \infty$) only depends on the tiling and the rate is higher for larger values of $r$ and $s$. In the previous section we established that there are infinitely many hyperbolic tilings subject to the constraint $1/r + 1/s < 1/2$. The tiling giving the smallest combined weight $r+s$ of the stabilizer generators is the \schl{5}{4}-tiling (and, of course, its dual \schl{4}{5}).
Compare the above result to the rate of codes obtained from tilings of oriented quotient surfaces of the euclidean plane. As discussed in the previous section they are topologically all tori. Hence, they only encode two logical qubits, regardless of the number of physical qubits.

What about the distance of hyperbolic surface codes? Unfortunately, there exists no formula for the combinatorial systole of a tiled quotient surface.

It has been proven in \cite{systole_bound} that for a \schl{r}{s}-tiling of $\hyp$ for any given $q\in \mathbb{N}$ there exists a hyperbolic surface  $\mathbb{H}^{2}/H$ tiled by $\lbrace r,s \rbrace$ with combinatorial systole larger than $q$ and $n\leq C^{q}s/2$, where $n$ is the number of edges and
\begin{align}\label{eqn:constant_distance}
  C\leq \frac{1}{2} (5rs)^{16rs} .
\end{align}
This means that there exists a family of quantum codes where the distance is lower bounded by a function growing logarithmically in the number of physical qubits (see also \cite{freedman2002z2}).
Since the logarithm is base $C$ and the value of $C$ is very large\footnote{There exist a follow-up result \cite{systole_better_bound} which involves number theoretic functions and gives a $C$ of some magnitudes smaller.}, this bound is not relevant for any practical purposes.

Unfortunately, we cannot do better than a logarithmic growth of the distance. In \cite{systole_lower_bound} the author proves that for a \schl{r}{s}-tiled quotient surface the combinatorial systole $d$ has an upper bound
\begin{align}
  d \leq \frac{r}{2} \log_{\sqrt{rs}}(2 n).
\end{align}
One can also compare this to the bound in Eqn. \ref{eqn:tradeoff}.

We can mention that hyperbolic surface codes of type \schl{3}{\infty}, where infinitely many triangles meet at so-called idealized vertices at infinity, have been studied in \cite{riemann}. However, these surfaces have a dual systole $\csys^{*}\in O(1)$, hence a distance upper bounded by a constant \cite{spini}.

Once a hyperbolic code is constructed, we can efficiently compute its distance using the following algorithm due to Bravyi \cite{bravyi_algo}:
First we compute a basis of the cocycle space $Z^{1}=\ker(\delta_{1})$ via Gau\ss ian elimination on the coboundary operator $\delta_{1}$, which takes $O(n^{3})$ steps where $n$ is the number of edges. Remember that the cocycles correspond to the logical $X$ operators (see Fig. \ref{fig:stabHomCorrespondence}).

Second, we consider two copies of the lattice $X$ and $X'$.\footnote{Since we are only interested in paths on the lattice we can consider them to be graphs and forget about the faces in either lattice.} For every element $b$ in the basis of $Z^{1}$ we define a new lattice $D$ using the following procedure: If the edge $e=(u,v)$ in $X$ and  $e'=(u',v')$ in $X'$ are in the support of $b$ they are removed from $X$ and $X'$. At the same time we introduce new edges connecting $u$ in $X$ with $v'$ in $X'$ and vice versa $u'$ with $v$. We basically take the original lattice $X$ and paste in a copy $X'$ along the coboundary $b$ to obtain $D$.
A non-trivial cycle has to have odd overlap with a cocycle, as established in Section \ref{sub:quantum_codes}. Hence, the length of the shortest non-trivial cycle in $X$  which intersects $b$ and passes through point $v$ can be determined by computing the shortest path between $v$ and $v'$ in $D$. This can be done using, for example, Dijkstra's algorithm which takes $O(n \log(n))$ steps. By looping over the $k$ elements of the cocycle basis $b$ and base points $v$ we can thus determine the systole using $O(n^3 + kn^{2}\log(n))$ steps. The cosystole is determined by repeating this procedure using the dual lattice.

\section{Constructions} %
\label{sec:constructions}

\subsection{Examples of Hyperbolic Surface Codes} %
\label{sub:examples}

In Section \ref{sub:quotient_surfaces} we established a correspondence between \schl{r}{s}-tiled, oriented quotient surfaces and fixed-point free, finite index, normal subgroups of the group $G_{r,s}^{+}$. In \cite{low_index_subgroups} the authors present an algorithm that, given the presentation of $G_{r,s}^{+}$ (Eqn.~\ref{eqn:gamma_plus}), finds {\rm all} of its normal subgroups $H$ up to a given index.

In Table \ref{tab:5_4_code_examples} and Table \ref{tab:8_3_code_examples} we present some examples of possible compactifications. The first three columns contain the three code parameters, followed by the combinatorial systole of the primal and the dual lattices.
The last column contains translations which compactify the lattice. In the group theoretic language of Section \ref{sub:quotient_surfaces} the translations and all conjugations by elements of $G_{r,s}$ generate the group $H$. If the compactification is achieved by  a single translation, then the length of this translation determines the combinatorial systole.

\begin{table*}
\begin{center}
\begin{tabular}{|c|c|c|c|c|c|}
\hline 
$n$ & $k$ & $d$ & $\csys$ & $\csys^{*}$ & translation(s) \\ 
\hline 
60 & 8 & 4 & 6 & 4 & $((\sigma\rho^{-1})^{2}\rho^{-1})^{2}$
 \\ 
\hline 
160 & 18 & 6 & 8 & 6 & $\sigma\rho^{2}(\sigma\rho^{-1})^{2}\rho^{-1}\sigma^{-2}\rho^{-2}\sigma\rho^{-1}
$ \\ 
\hline 
360 & 38 & 8 & 8 & 8 & $(\sigma\rho^{2}\sigma)^{2}(\rho^{-1}\sigma^{-2}\rho^{-1})^{2}
$ \\ 
\hline 
1800 & 182 & 10 & 10 & 10 & $(\sigma\rho^{-1})^{10}, \sigma\rho^{2}\sigma^{2}\rho^{-1}\sigma(\rho^{2}\sigma^{-1})^{2}(\rho\sigma^{-1})^{2}\sigma^{-1}\rho^{-2}\sigma\rho^{-1}$ \\ 
\hline 
1920 & 194 & 10 & 12 & 10 & $\sigma\rho^{2}\sigma^{2}\rho(\rho\sigma^{-1})^{4}\rho^{-1}(\rho^{-1}\sigma)^{3}\rho^{-1}$ \\ 
\hline 
\end{tabular} 
\end{center}
	\caption{Examples of compactified \schl{5}{4}-lattices. The examples were chosen such that we have a sequence with growing $n$ and $d$. The second to last lattice is compactified by two independent translations.} \label{tab:5_4_code_examples}
\end{table*}

\begin{table}
\begin{center}
\begin{tabular}{|c|c|c|c|c|c|}
\hline 
$n$ & $k$ & $d$ & $\csys$ & $\csys^{*}$ & translation \\ 
\hline 
48 & 4 & 3 & 6 & 3 & $(\rho^{2}\sigma^{-1})^{3}
$ \\ 
\hline 
168 & 14 & 4 & 8 & 4 & $(\sigma\rho^{-2})^{4}
$ \\ 
\hline 
384 & 32 & 4 & 12 & 4 & $(\sigma\rho^{-3})^{4}
$ \\ 
\hline 
648 & 54 & 6 & 14 & 6 & $\sigma\rho^{4}\sigma\rho^{-1}\sigma\rho^{2}\sigma^{-1}\rho^{3}\sigma^{-1}\rho^{-3}\sigma\rho^{-1}$ \\ 
\hline 
768 & 64 & 6 & 16 & 6 & $\sigma\rho^{2}(\rho^{2}\sigma^{-1})^{3}\rho^{3}\sigma^{-1}\rho^{-3}\sigma\rho^{-2}$ \\ 
\hline 
\end{tabular} 
\end{center}
\caption{Examples of compactified \schl{8}{3}-lattices. The combinatorial systole of the primal lattices is larger than that of the dual lattices.} \label{tab:8_3_code_examples}
\end{table}

\subsection{Numerical Analysis of Codes} %
\label{sub:numerical_analysis}

In this section we present the results of simulated error corrections on hyperbolic surface codes. Our error model is that of independent $X$ and $Z$ errors in which a qubit can undergo independently an $X$ error with probability $p$ {\em and} a $Z$ error with probability $p$ at each time-step.
After these errors happen one applies the MWM-decoder which tries to infer what error occurred. Hereby we assume that syndrome measurements can be done perfectly. 
The decoder succeeds if the product of the real and the inferred error is in the image of the boundary operator. We do this independently for the dual lattice as well.

We gather statistics by repeating this procedure $N$ times. The probability of a logical error $P_{log}$ on \emph{any} of the encoded qubits is estimated by taking the ratio between failed corrections and number of trials. In the usual simulation of the toric code one plots the logical error rate against one type of error ($X$ or $Z$) only, as the quantum error correction for both types of errors is identical. This is not the case for general \schl{r}{s} codes.

\begin{figure}
    \centering
    \begin{subfigure}[b]{0.45\textwidth}
        \includegraphics[width=\linewidth]{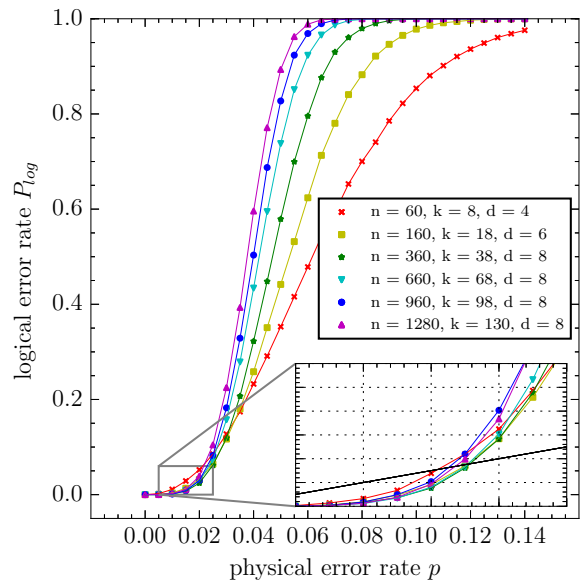}
        \caption{The zoomed-in plot shows $p$ in the interval $[0.005, 0.025]$ and $P_{log}$ in the interval $[0, 0.06]$. The black line in the zoomed-in plot represents $p=P_{log}$.}
        \label{fig:54zoomed}
    \end{subfigure}
    	~ 
    \begin{subfigure}[b]{0.45\textwidth}
        \includegraphics[width=\linewidth]{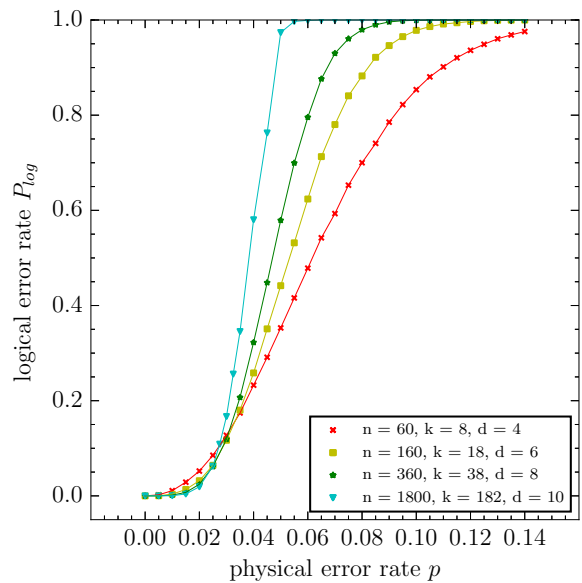}
        \caption{$P_{log}$-$p$-diagram for \schl{5}{4}-tiled surfaces. The examples were chosen such that the ratio of distance to physical qubits forms a strictly increasing sequence.}
        \label{fig:54optimal}
    \end{subfigure}
  \caption{Logical error $P_{log}$ against physical error $p$ for \schl{5}{4}-tiled surfaces. Every data point was obtained from $4\times 10^{4}$ runs.}
  \label{fig:P_p_5_4}
\end{figure}

The results of a simulation with $N=4\times 10^4$ trials are shown in Figure \ref{fig:P_p_5_4}. For increasing probability of a physical error $p$ the probability for a logical error occuring on any of the qubits $P_{log}$ approaches $1-(1/2)^{2k}$. This is because the decoder can do no worse than random guessing and the probability to guess correctly for all $2k$ logical operators $\overline{X}_i,\overline{Z}_i$ is $(1/2)^{2k}$ (for the toric code this is $0.75$ for a single type of Pauli error).

In a numerical simulation of the noise threshold of the toric code (see e.g. \cite{toric_threshold}) the logical error probability curves for different numbers of qubits $n$ all intersect at one point, namely the threshold $p_{c}$, clearly pointing to the fact that the noise threshold corresponds to a phase-transition.
In Fig.~\ref{fig:54zoomed} we see that the distributions shift to the left with increasing number of qubits. Note that even if the distance is not increasing we can still observe a better protection against errors. We believe that this is the case due to an increase in the average minimum weight of the operators of the logical qubits.

When taking a sequence of \schl{5}{4}-codes with strictly increasing distance and number of physical qubits, we see that the first three lines cross (Fig.~\ref{fig:54optimal}). However, the line corresponding to the \schl{5}{4}-code with $n=1800$ qubits does not cross. The data is somewhat inconclusive in showing whether there is a noise threshold and what its asymptotic value is. Further examples can be found in the Appendix \ref{sec:appendix}.
It is important to note that it has been shown in \cite{thresholds_proof} that there exists a finite threshold for all homological codes with distance lower bounded by a logarithm in the number of qubits, when quantum error correction is perfect or noisy.

For hyperbolic surface codes, we expect the existence of another threshold where all logical qubits are potentially corrupted. In \cite{perculation} the authors analyze the two distinct perculation thresholds for a prototype model of a hyperbolic lattice. Here we only consider the threshold where an arbitrary qubit becomes corrupted.

Our data suggests that for the \schl{5}{4}-tiling there exists a threshold between 1\% and 5\% (see Figure \ref{fig:P_p_5_4}). For the toric code the threshold against this error model is $p_c=10.3\%$ as determined in \cite{toric_threshold}.
 
The distance of the code depends on the parameters $r$ and $s$ of the tiling. For increasing $r,s$ (and thus a code with better rate) the lower bound on the distance becomes smaller and one would expect that the threshold also goes down (at least the lower bound on the threshold in \cite{thresholds_proof} becomes smaller).

In our simulations of various $\{r,s\}$ codes we see that the threshold does go down for increasing stabilizer weight $r$ and $s$ which is conform to our expectation (see Figure \ref{fig:P_p_high} in Appendix \ref{sec:appendix}).

\begin{figure}[ht]
  \begin{center}    
    \includegraphics[width=\columnwidth]{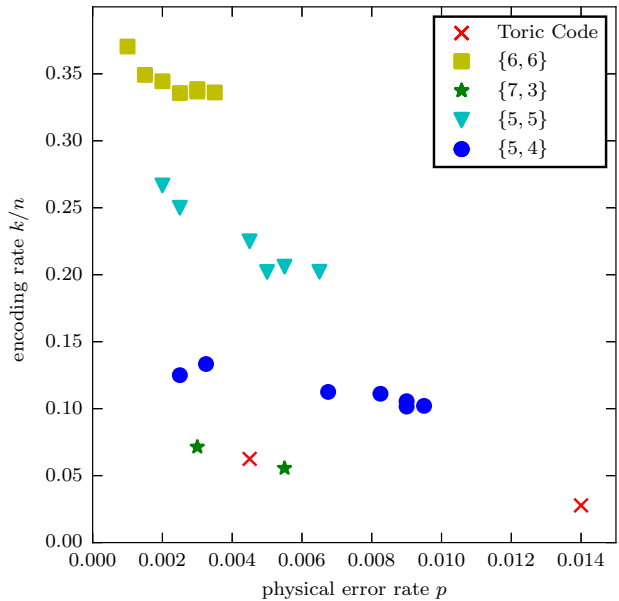}
  \end{center}
  \caption{Encoding rate of a code which protects qubits with probability $>0.999$. The number of physical qubits varies between 60 and 960. Data points are labeled by the tiling. The two instances of the toric code are $L=4$ and $L=6$.}
  \label{fig:overhead}
\end{figure}

To compare the toric code to the hyperbolic surface codes we consider the overhead to produce logical qubits and protect them against decoherence. To do this we mark up to which physical error rate a code can protect all of its qubits with probability at least 0.999. The results are shown in Figure \ref{fig:overhead}.

We see that the toric codes protect against errors up to a higher physical error rate. However, their overhead is quite big compared to the hyperbolic surface codes.

For each family of codes the encoding rate decreases with $n$ (Eqn.~\ref{eqn:hyp_rate}). Since we are below threshold and $P_{log}$ goes to zero for increasing $n$ the capability of the codes to protect against errors increases. For each code family we obtain a slope which converges to the asymptotic rate of the code. Codes with lower rate offer better protection against errors. Since the weight of the logicals is not a monotone function in the number of physical qubits the slopes are not monotone either.

The hyperbolic surface code with the smallest encoding rate is the \schl{7}{3}-code which was analyzed in \cite{isaac}. There are only two \schl{7}{3} codes with less than $10^{3}$ qubits. The next smallest rate is achieved by the \schl{5}{4}-code which appears to offer the best error protection of all hyperbolic surface codes.

\begin{table}[ht]
\centering
\begin{tabular}{|c|c|c|c|l|l|}
  \hline
Lattice     & $n$  & $k$ &  $d$  &   weight of $Z$-logicals   & weight of $X$-logicals     \\ \hline
\schl{5}{4} & 60 & 8 &  4  & 6 (8)                   & 4 (8)                           \\ \hline 
\schl{5}{5} & 30 & 8 & 3  & 3 (6), 4 (2)          & 3 (8)                          \\
           & 40 & 10 & 4 & 4 (10)                 & 4 (9), 5, (1)                  \\
\hline 
\schl{6}{4} & 36 & 8 & 4  & 4 (4), 6 (4)          & 4 (8)                          \\
\hline 
\schl{6}{5} & 60 & 18 & 3 & 4 (15), 6 (3)         & 3, (3), 4, (13), 5 (1), 6 (1) \\
\hline 
\schl{6}{6} & 54 & 20 & 4 & 4 (18), 6 (2)         & 4 (20)                         \\
             & 60 &  22 & 4  & 4 (15), 5 (6), 6 (1) & 4 (16), 5 (6)           \\      \hline                           
\end{tabular}
\caption{Minimum weight of the logical operators of some small hyperbolic surface codes. The number in parenthesis is the number of distinct logical operators with the same minimum weight.}
\label{table:distances}
\end{table}

In Table \ref{table:distances} we show the minimum weight of the logical operators of some hyperbolic surface codes. The weights of the logicals were determined by brute force as the method to determine the distance from Section \ref{sub:properties_of_hyperbolic_surfaces} does not directly apply.

The smallest instance of the \schl{5}{4}-code encodes $k=8$ logical qubits into $n=60$ physical qubits and has distance $d=4$. An efficient use of the surface code (in which the lattice is chopped off at the boundaries, see \cite{horsman_surface}) has parameters $[[n=d^2,k=1,d]]$ (instead of $[[n=d^2+(d-1)^2,k=1,d]]$). Hence to encode 8 qubits with the surface code with distance $4$ requires 128 physical qubits, showing that the hyperbolic construction can lead to more efficient coding than the surface code.

Note that the surface code has a higher threshold, making it more favourable for noisy qubits. For more coherent qubits hyperbolic surface codes might offer an advantage.

\subsection{Hyperbolic surface codes with open boundaries} %
\label{sub:planar_hyperbolic_codes}

One can try to construct a hyperbolic surface code with open boundaries in various ways. Such codes, if they have good distance and rate, would have a larger practical appeal than the hyperbolic surface codes, as they would correspond to planar graphs.

\begin{figure}[ht]
  \begin{center}
   \includegraphics[width=0.8\columnwidth]{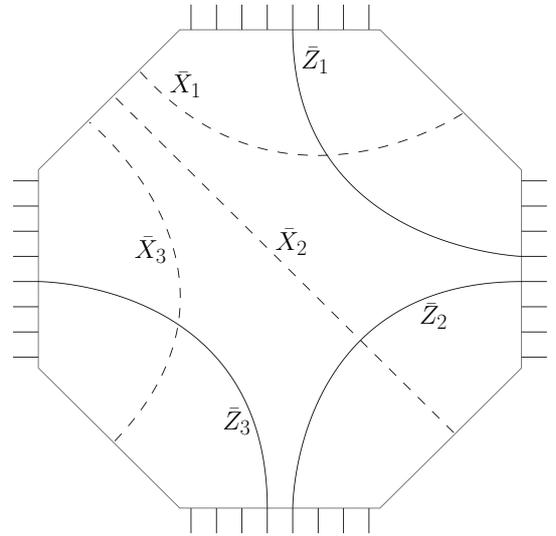} 
  \end{center}
  \caption{A polygon with $2k$ sides, alternatingly `rough' and `smooth' (see definitions in Section \ref{sub:small_planar_code_example}), can encode $k-1$ logical qubits \cite{surface_code}. Shown is $k=4$ and a choice for the  logical $\overline{X}_i, \overline{Z}_i$ operators. }\label{fig:kgon}
\end{figure}

A first idea is to cut open the multi-handled torus to obtain a surface with some boundaries. One method produces a surface with punctures, the other method would produce a region as depicted in Fig.~\ref{fig:kgon}. 

In \cite{freedman_projective} it was first described how a surface with punctures can encode multiple qubits. This type of encoding is used in the surface code architecture described in \cite{fowler:practical} where qubits are encoded in both rough or smooth double-holes punctured in the surface. For this encoding which uses both smooth and rough holes, there is a simple scheme with which one can perform a CNOT gate, first described in \cite{RH:cluster2D}. 

One could image obtaining such a punctured surface by cutting open the multi-handled torus. One would start with a torus with $g$ handles and no boundary, nor punctures, encoding $k=2g$ qubits. If one cuts open one handle, one has two punctures (and $g-1$ handles) in the remaining surface. Cutting open another disjoint handle similarly generates two punctures and cutting open the last handle leads to a surface with $2g-1$ punctures and one outer boundary (the last puncture). For a tiled surface one can make a smooth cut or a rough cut.
For a smooth cut, one follows a non-trivial $\overline{Z}$-loop $\gamma$ on the graph around the handle and all edges $e\in E_{\gamma}$ on this loop $\gamma$ are replaced by double pairs of edges, i.e. $E_{\gamma} \rightarrow E_{\gamma}^A \cup E_{\gamma}^B$ so that the two $Z$-plaquettes adjoining these edges either act on edges in $E_{\gamma}^A$ or $E_{\gamma}^B$ (but not both). At each vertex on this path the number of $X$-check operators is doubled such that one $X$-check operator acts on the edges in $E_{\gamma}^A$ and the other on the edges in $E_{\gamma}^B$. Note that in this procedure, one adds as many qubits as one adds $X$-check operators as the number of edges and vertices around a loop are the same. For the many-handled torus, there is one linear dependency between all plaquette $Z$-checks (the product of all them is $I$) and similarly one linear dependency for the $X$-checks. For the final punctured surface (with smooth holes), there is only a linear dependency between all $X$-checks:
this directly implies that the punctured surface will encode $2g-1$ logical qubits.  Another way of seeing that a surface with a smooth outer boundary and $2g-1$ smooth punctures or holes encodes $2g-1$ logical qubits is by enumerating the logical operators: for each hole the logical $\overline{Z}$ is a $Z$-loop around the hole, while the logical $\overline{X}$ is a $X$-loop on the dual lattice to the outer boundary (see e.g. \cite{quantumcodes_review}).
However, one does not need to use all these logical qubits. 

In the smooth double-hole encoding described in \cite{fowler:practical, quantumcodes_review}, one uses a pair of holes to encode one qubit where the logical $\overline{Z}$ is the $Z$-loop around any of the two holes and the logical $\overline{X}$ is the $X$-distance between the two holes. In this way, the surface with $2g-1$ holes can encode $g$ double-hole qubits where the last double-hole qubit is formed by the last hole plus the boundary `hole'.  \\

However, does this procedure preserve the distance of the original code and is it even possible to execute this procedure on the graph obtained from a \schl{r}{s}-tiling of the closed hyperbolic surface?
The answer is in fact no. If we have a family of $\{r,s\}$ codes with increasing $k \sim n$ and distance lower bounded as some $c(r,s) \log n$, then logical $\overline{Z}$ loops (which act on at least on $c(r,s) \log n$ qubits) must overlap on many qubits. If there are $\Omega(k)$ such non-intersecting loops, then there must be at least $\Omega(k \log n)=\Omega(n \log n)$ qubits in total, which is not the case! Hence, one cannot find such a set of non-overlapping loops along which to cut as logical operators must share a lot of space (support).

Another method of cutting the many-handled tiled torus to create smooth and rough boundaries would allow one to create a coding region as in Fig.~\ref{fig:kgon}.  One can encode multiple qubits into such a surface code \cite{surface_code} by using alternating rough and smooth boundary of sufficiently large length. The boundary of the encoding region is divided into $2k$ regions and encodes $k-1$ logical qubits.
The logical $\overline{Z}$ operators can start and terminate at rough boundary regions while the logical $\overline{X}$ operators start and terminate at smooth boundary regions. One can consider the asymptotic scaling of the parameters $n,k$ and $d$ of the code represented by such an encoding region, where we do not assume anything about the distribution of qubits inside the region, meaning that for such a class of codes with increasing $n$ qubits can get closer together (or further apart) if we represent the qubits on the euclidean plane. One can simply argue that for such a family of homological surface codes with boundaries the following bound should hold
\begin{equation}
k d\leq  c n,
\label{eq:bound_kdn}
\end{equation}
with a constant $c$. This bound shows that for codes based on tiling a surface with open boundaries, it is not possible to have a constant encoding rate $k/n$ {\em and} a distance increasing as $\log n$. We argue this bound for the encoding region in Fig.~\ref{fig:kgon}, but similar arguments should hold for the tiling of a surface with holes. The simple argument goes as follows. In order to encode $k-1$ qubits, one divides the set of boundary edges $E_{\rm bound}$ into $2k$ regions. Clearly, the total number of qubits $n \geq E_{\rm bound}$. The logical operators run from rough to a rough or smooth to smooth boundary, hence their distance is upper-bounded by the length of a region $c |E_{\rm bound}|/k \leq c n/k$ for some constant $c$. This is true as one can always let the logical operator run along (or close to) the boundary edges. This results in Eqn.~\ref{eq:bound_kdn}.
If the encoding region is a disk with many holes, we can argue similarly.  We enumerate the number of edges (qubits) around the non-trivial disjoint holes which should be less than the total number of qubits $n$. The number of logical qubits $k$ scales linearly with the number of holes and the distance $d$ is at most the number of edges around each hole, resulting in Eq.~\ref{eq:bound_kdn}.

Thus we may ask whether using hyperbolic geometry gives any advantage over euclidean geometry when considering tiled surfaces with a boundary. The previously proven bound $kd^2 \leq cn $ is still worse than the simple bound $k d \leq cn$. It is worth noting that encoding via the boundaries as in Fig.~\ref{fig:kgon} has worse scaling than an optimal surface code encoding in the case of a euclidean metric on the underlying qubits. In this case, the perimeter of the polygon scales at most as $\sqrt{n}$ where $n$ is the total number of qubits. This implies that the distance of the code $d \leq c \sqrt{n}/k$, which is worse than the $k d^2 \leq c n$ bound. If we use hyperbolic geometry and imagine the encoding region as a partially-tiled hyperbolic plane, see the explicit construction below, then the number of qubits at the boundary scales like the total number of qubits, which seems promising. However, the minimum weight logical operators which run from boundary to boundary, will run along shortest paths, geodesics, which go through the interior of the region. We show a small explicit example of such a code in the next section, but the underlying construction is unlikely to give good asymptotic scaling behavior.

\subsection{Small Planar Code Example}\label{sub:small_planar_code_example}

\begin{figure}[ht]
	  \begin{center}
  \includegraphics[width=0.8\columnwidth]{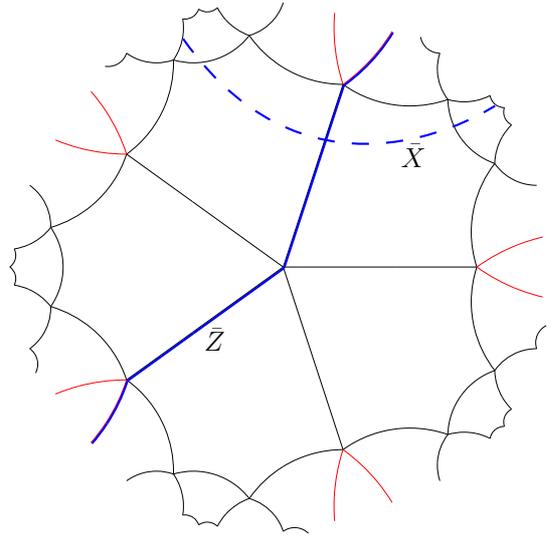} 
  \end{center}
	\caption{A $[[65,4,4]]$ code based on the \{5,5\} tiling. The distance of the logical $\overline{X}$ is in fact 5 while it is 4 for the logical $\overline{Z}$. The number of boundary edges in $E_{\rm bound}$ of the starting graph was 60 and was  divided into 10 regions each with 6 edges. Shown in red are the additional weight-2 $ZZ$ checks.} \label{fig:55partial_tile}
\end{figure}

We generate a planar graph with a finite number of vertices by taking a regular tiling $\{r,s\}$, partially tiling the hyperbolic plane and then modifying the boundary so that the planar graph encodes logical qubits, as in Fig.~\ref{fig:kgon}.  An example of the procedure to create such a graph encoding multiple qubits consists of the following steps; we illustrate the idea in Fig.~\ref{fig:55partial_tile} for the $\{5,5\}$ tiling:

1. For a $\{r,s\}$ tiling, start with a single $r$-gon (call it level-1 $r$-gon) and reflect in the edges of this $r$-gon to generate level-2 $r$-gons. Repeat this for the level-2 $r$-gons etc. so that one obtains a planar graph $G=(V,E)$ where the faces are the level-1 to level-m $r$-gons for some $m$. In the Figure, we have started with 4 level-1 $r$-gons and generated only the level-2 $r$-gons and have stopped there.
With every face $f$ we associate a plaquette $Z$-check acting on the qubits on the boundary of this face, and with every vertex $v$ of this graph one associates a star $X$-check acting on the qubits on edges adjacent to the vertex. At the boundaries the weight of the $X$-checks will be two or $s$ (for even $s$) or $s-1$ (for odd $s$), in the interior the $X$-checks have weight $s$. The code associated with this {\em starting graph} $G$ encodes no qubits.
There is one linear dependency between all $X$-checks and so the number of linearly independent $X$-checks is $V-1$. There is no linear dependency between the $Z$-checks of which there are $F$. As this is a planar graph (with Euler characteristic $\chi=1$), one has $E=V-1+F$ where $E$ is the number of edges, equal to the number of qubits $n$, hence no encoded qubits. \\

All boundaries in this graph are so-called smooth boundaries at which a string of $X$-errors can start and end. For this graph which encodes no qubits, such a string can be annihilated by star $X$-checks, hence there are no logical operators. Thus we need to modify this graph in order to create so-called rough boundaries. At rough boundaries a string of $Z$-errors can start. If one alternates rough and smooth boundaries as in Fig.~\ref{fig:kgon}, these strings can no longer be annihilated by $X$- or $Z$-checks, but have to run from boundary to boundary. What is the procedure for creating several rough boundaries at which $Z$-strings can end? \\

2. First, for the given starting graph $G$, one counts the number of boundary edges in $E_{\rm bound}$ as $|E_{\rm bound}|$: these are defined to be the edges that the level-$k$ $r$-gons would be reflected in to generate level $k+1$ $r$-gons. In order to create a code where all encoded qubits have about the same distance, one wants to divide the set $E_{\rm bound}$ in $2k$ equally-sized sets or regions with $k > 2$ so that each subset of size $\lfloor |E_{\rm bound}|/(2k)\rfloor$ corresponds to a rough or smooth boundary. Such code can encode $k-1$ qubits \cite{surface_code}: the logical $\overline{Z}_i$, $i=1,\ldots, k-1$ will run from the $i$th to the $(i+1)$th rough boundary, while the logical $\overline{X}_i$ will run from the smooth boundary in between the $i$th and $(i+1)$th rough boundary to the $(i+1)$th smooth boundary (see Fig.~\ref{fig:kgon}). If $k$ is chosen too large, then each smooth/rough region would be too short and the shortest path between two regions would not go through the bulk. Hence one should choose $k$ such that the shortest path along the perimeter is about the same length as the shortest path through the bulk: in that case one uses the available qubit-space optimally. This choice is illustrated in Fig.~\ref{fig:55partial_tile} where one has 60 edges along the boundary which we divide up into $10=2k$ regions of $6$ edges each, thus encoding $4$ logical qubits.

The creation of a rough boundary in a region of edges $E_{\rm bound}^{\rm region}\subseteq E_{\rm bound}$ consists of the following 3 steps (variants are possible). First, one removes all $X$-check operators of weight-2 which have 2 edges among $E_{\rm bound}^{\rm region}$ (if the weight-2 $X$-check has an edge in $E_{\rm bound}^{\rm region}$ one also removes it).  
One could in principle also remove checks with weight more than 2 at the boundary but this makes the lattice a bit smaller, so in this construction we prefer to remove only the weight-2 checks. Then all qubits on which only a single plaquette $Z$-check acts are removed from the lattice and these plaquettes are thus modified (in the toric code they are the weight-3 plaquettes at the boundary). The removal of the $X$-checks makes it possible for a $Z$-string to start on an edge attached to a removed $X$-check. We need to make sure that such a string can only run from rough boundary to another rough boundary. Since we did not remove some $X$-checks in the rough region (some weight-5 checks in Fig.\ref{fig:55partial_tile}), in general of weight $s$), one needs to add weight-2 $ZZ$ checks to the stabilizer, as these operators commute with all the current checks. These weight-2 checks are indicated (in red) in Fig.~\ref{fig:55partial_tile}.
For the construction in Fig.~\ref{fig:55partial_tile}, one can verify that the minimum weight logical $\overline{Z}$ is of weight-4, while the minimum-weight logical $\overline{X}$ is of weight-5: we draw examples of these logical operators in the Figure. The total number of physical qubits is $n=65$. 
The most efficient use of the surface code (in which the lattice is chopped off at the boundaries, see \cite{horsman_surface}) has parameters $[[d^2,1,d]]$ (instead of $[[d^2+(d-1)^2,1,d]]$). Hence to encode 4 qubits with the surface code with distance $4$ requires 64 physical qubits and distance $5$ would require 100 qubits, showing that this simple hyperbolic construction can lead to a somewhat more efficient coding than the surface code. \\

In order to construct a planar graph encoding $k-1$ logical qubits, one can also generate the planar starting graph $G$ encoding no qubits by repeatedly rotating an elementary $r$-gon by $2\pi/s$ around its vertices. By this rotation, one generates a new generation of $r$-gons, the level-2 $r$-gons etc. 
In this construction, one removes all $X$-checks in the region where the rough boundary has to be formed. This means that no weight-2 $ZZ$ checks need to be added. Only vertices on which a single plaquette $Z$-operator acts are to be further removed.  \\

\section{Conclusion} %
\label{sec:conclusion}

We have discussed explicit constructions of families of 2D topological quantum codes derived from regular tilings on surfaces with negative curvature. On a hyperbolic surface without boundaries these hyperbolic surface codes have an asymptotically constant rate and low $O(1)$ weight stabilizer checks and distance scaling as $\log n$ for a total of $n$ physical qubits. This trade-off of distance versus encoding rate is much more favorable than for the toric or surface code. For practical purposes it is desirable to minimize the weight of the check operators as the quantum circuit through which one measures the check operators is noisy. For this reason, the most attractive code may be the $\{5,4\}$ hyperbolic surface code with an asymptotic rate of $k/n \rightarrow 1/10$: we have obtained some partial evidence that the noise threshold of this code for noiseless error correction is above $2\%$. 
It may of be interest to consider the explicit usage of these codes for a 2D quantum memory given that they allow for a higher-qubit storage density.

\section*{Acknowledgments} %
\label{sec:acknowledgment}
We would like to thank David DiVincenzo for interesting discussions. We acknowledge funding through the EU program QALGO. NPB would like to thank Jascha Ulrich for his help with \textsc{matplotlib}.

\bibliographystyle{IEEEtran}
\bibliography{IEEEabrv,2d_homcodes_final}

\begin{thebibliography}{10}
\providecommand{\url}[1]{#1}
\csname url@samestyle\endcsname
\providecommand{\newblock}{\relax}
\providecommand{\bibinfo}[2]{#2}
\providecommand{\BIBentrySTDinterwordspacing}{\spaceskip=0pt\relax}
\providecommand{\BIBentryALTinterwordstretchfactor}{4}
\providecommand{\BIBentryALTinterwordspacing}{\spaceskip=\fontdimen2\font plus
\BIBentryALTinterwordstretchfactor\fontdimen3\font minus
  \fontdimen4\font\relax}
\providecommand{\BIBforeignlanguage}[2]{{%
\expandafter\ifx\csname l@#1\endcsname\relax
\typeout{** WARNING: IEEEtran.bst: No hyphenation pattern has been}%
\typeout{** loaded for the language `#1'. Using the pattern for}%
\typeout{** the default language instead.}%
\else
\language=\csname l@#1\endcsname
\fi
#2}}
\providecommand{\BIBdecl}{\relax}
\BIBdecl

\bibitem{quantumcodes_review}
\BIBentryALTinterwordspacing
B.~M. Terhal, ``Quantum error correction for quantum memories,'' \emph{Rev.
  Mod. Phys.}, vol.~87, pp. 307--346, Apr 2015. [Online]. Available:
  \url{http://link.aps.org/doi/10.1103/RevModPhys.87.307}
\BIBentrySTDinterwordspacing

\bibitem{kitaev:survey}
A.~Y. Kitaev, ``Quantum computations: algorithms and error correction,''
  \emph{Russian Math. Surveys}, vol.~52, pp. 1191--1249, 1997.

\bibitem{surface_code}
S.~B. Bravyi and A.~Y. Kitaev, ``Quantum codes on a lattice with boundary,''
  \emph{arXiv preprint quant-ph/9811052}, 1998.

\bibitem{freedman_projective}
M.~H. Freedman and D.~A. Meyer, ``Projective plane and planar quantum codes,''
  \emph{Foundations of Computational Mathematics}, vol.~1, no.~3, pp. 325--332,
  2001.

\bibitem{toric_code}
E.~Dennis, A.~Kitaev, A.~Landahl, and J.~Preskill, ``Topological quantum
  memory,'' \emph{Journal of Mathematical Physics}, vol.~43, no.~9, pp.
  4452--4505, 2002.

\bibitem{Kelly2015}
\BIBentryALTinterwordspacing
J.~Kelly \emph{et~al.}, ``State preservation by repetitive error detection in a
  superconducting quantum circuit,'' \emph{Nature}, vol. 519, no. 7541, pp.
  66--69, mar 2015. [Online]. Available:
  \url{http://dx.doi.org/10.1038/nature14270}
\BIBentrySTDinterwordspacing

\bibitem{Crcoles2015}
\BIBentryALTinterwordspacing
A.~C{\'{o}}rcoles, E.~Magesan, S.~J. Srinivasan, A.~W. Cross, M.~Steffen, J.~M.
  Gambetta, and J.~M. Chow, ``Demonstration of a quantum error detection code
  using a square lattice of four superconducting qubits,'' \emph{Nature
  Communications}, vol.~6, p. 6979, apr 2015. [Online]. Available:
  \url{http://dx.doi.org/10.1038/ncomms7979}
\BIBentrySTDinterwordspacing

\bibitem{Rist2015}
\BIBentryALTinterwordspacing
D.~Rist{\`{e}}, S.~Poletto, M.-Z. Huang, A.~Bruno, V.~Vesterinen, O.-P. Saira,
  and L.~DiCarlo, ``Detecting bit-flip errors in a logical qubit using
  stabilizer measurements,'' \emph{Nature Communications}, vol.~6, p. 6983, apr
  2015. [Online]. Available: \url{http://dx.doi.org/10.1038/ncomms7983}
\BIBentrySTDinterwordspacing

\bibitem{no_go_thm}
S.~Bravyi and B.~Terhal, ``A no-go theorem for a two-dimensional
  self-correcting quantum memory based on stabilizer codes,'' \emph{New Journal
  of Physics}, vol.~11, no.~4, p. 043029, 2009.

\bibitem{BPT_tradeoffs}
S.~{Bravyi}, D.~{Poulin}, and B.~{Terhal}, ``{Tradeoffs for Reliable Quantum
  Information Storage in 2D Systems},'' \emph{Physical Review Letters}, vol.
  104, no.~5, p. 050503, Feb. 2010.

\bibitem{fetaya_master_thesis}
E.~Fetaya, ``Homological error correcting codes and systolic geometry,'' 2011,
  {M}aster thesis, {E}instein {I}nstitute of {M}athematics {E}dmund {J}.
  {S}atillfra campus of the {H}ebrew {U}niversity.

\bibitem{freedman2002z2}
M.~H. Freedman, D.~A. Meyer, and F.~Luo, ``{$Z_2$}-systolic freedom and quantum
  codes,'' \emph{Mathematics of quantum computation, Chapman \& Hall/CRC}, pp.
  287--320, 2002.

\bibitem{delfosse_tradeoffs}
N.~Delfosse, ``Tradeoffs for reliable quantum information storage in surface
  codes and color codes,'' in \emph{Information Theory Proceedings (ISIT), 2013
  IEEE International Symposium on}.\hskip 1em plus 0.5em minus 0.4em\relax
  IEEE, 2013, pp. 917--921.

\bibitem{zemor2009cayley}
G.~Z{\'e}mor, ``On {C}ayley graphs, surface codes, and the limits of
  homological coding for quantum error correction,'' in \emph{Coding and
  cryptology}.\hskip 1em plus 0.5em minus 0.4em\relax Springer Berlin, 2009,
  pp. 259--273.

\bibitem{isaac}
I.~Kim, ``Quantum codes on {H}urwitz surfaces,'' 2007, {B}achelor thesis,
  {MIT}.

\bibitem{divincenzo_arch}
D.~P. {DiVincenzo}, ``{Fault-tolerant architectures for superconducting
  qubits},'' \emph{Physica Scripta Volume T}, vol. 137, no.~1, p. 014020, Dec.
  2009.

\bibitem{pastawski2015holographic}
F.~Pastawski, B.~Yoshida, D.~Harlow, and J.~Preskill, ``Holographic quantum
  error-correcting codes: Toy models for the bulk/boundary correspondence,''
  \emph{arXiv preprint arXiv:1503.06237}, 2015.

\bibitem{stillwell}
J.~Stillwell, \emph{Geometry of surfaces}, ser. Universitext.\hskip 1em plus
  0.5em minus 0.4em\relax New York, Berlin, Heidelberg: Springer-Verlag, 1992.

\bibitem{ratcliffe}
J.~Ratcliffe, \emph{{F}oundations of {H}yperbolic {M}anifolds}.\hskip 1em plus
  0.5em minus 0.4em\relax Springer New York, 2006, no. v. 10.

\bibitem{systole_bound}
M.~Ma{\v{c}}aj, J.~{\v{S}}ir{\'a}{\v{n}}, and M.~Ipolyiov{\'a}, ``Planar width
  of regular maps,'' \emph{Electronic Notes in Discrete Mathematics}, vol.~28,
  pp. 477--484, 2007.

\bibitem{systole_better_bound}
------, ``Injectivity radius of representations of triangle groups and planar
  width of regular hypermaps,'' \emph{Ars Mathematica Contemporanea}, vol.~1,
  no.~2, pp. 223--241, 2008.

\bibitem{systole_lower_bound}
J.~F. Moran, ``The growth rate and balance of homogeneous tilings in the
  hyperbolic plane,'' \emph{Discrete Mathematics}, vol. 173, no.~1, pp.
  151--186, 1997.

\bibitem{riemann}
R.~Riemann, ``Good families of quantum low-density parity-check codes and a
  geometric framework for the amplitude-damping channel,'' 2011, {Ph.D.
  thesis}, {MIT}.

\bibitem{spini}
G.~Spini, ``A study of quantum error-correcting codes derived from platonic
  tilings,'' 2013, {M}aster thesis, {I}nstitut de {M}athématiques de
  {B}ordeaux, {{U}niversit\'e de {B}ordeaux}.

\bibitem{bravyi_algo}
S.~Bravyi, {U}npublished.

\bibitem{low_index_subgroups}
M.~Conder and P.~Dobcs{\'a}nyi, ``Applications and adaptations of the low index
  subgroups procedure,'' \emph{Mathematics of computation}, vol.~74, no. 249,
  pp. 485--497, 2005.

\bibitem{toric_threshold}
C.~Wang, J.~Harrington, and J.~Preskill, ``Confinement-{H}iggs transition in a
  disordered gauge theory and the accuracy threshold for quantum memory,''
  \emph{Annals of Physics}, vol. 303, no.~1, pp. 31 -- 58, 2003.

\bibitem{thresholds_proof}
I.~Dumer, A.~A. Kovalev, and L.~P. Pryadko, ``Thresholds for correcting errors,
  erasures, and faulty syndrome measurements in degenerate quantum codes,''
  \emph{arXiv preprint arXiv:1412.6172}, 2014.

\bibitem{perculation}
\BIBentryALTinterwordspacing
P.~Minnhagen and S.~K. Baek, ``Analytic results for the percolation transitions
  of the enhanced binary tree,'' \emph{Phys. Rev. E}, vol.~82, p. 011113, Jul
  2010. [Online]. Available:
  \url{http://link.aps.org/doi/10.1103/PhysRevE.82.011113}
\BIBentrySTDinterwordspacing

\bibitem{horsman_surface}
C.~{Horsman}, A.~G. {Fowler}, S.~{Devitt}, and R.~{Van Meter}, ``{Surface code
  quantum computing by lattice surgery},'' \emph{New Journal of Physics},
  vol.~14, no.~12, p. 123011, Dec. 2012.

\bibitem{fowler:practical}
A.~G. {Fowler}, M.~{Mariantoni}, J.~M. {Martinis}, and A.~N. {Cleland},
  ``{Surface codes: Towards practical large-scale quantum computation},''
  \emph{Phys. Rev. A}, vol.~86, no.~3, p. 032324, Sep. 2012.

\bibitem{RH:cluster2D}
R.~{Raussendorf} and J.~{Harrington}, ``{Fault-tolerant quantum computation
  with high threshold in two dimensions},'' \emph{Phys. Rev. Lett.}, vol.~98,
  no.~19, p. 190504, May 2007.

\end{thebibliography}

\newpage

\appendix\label{sec:appendix}

\begin{figure}[ht]
  \centering
    \begin{subfigure}{.43\textwidth}
      \centering
      \includegraphics[width=\linewidth]{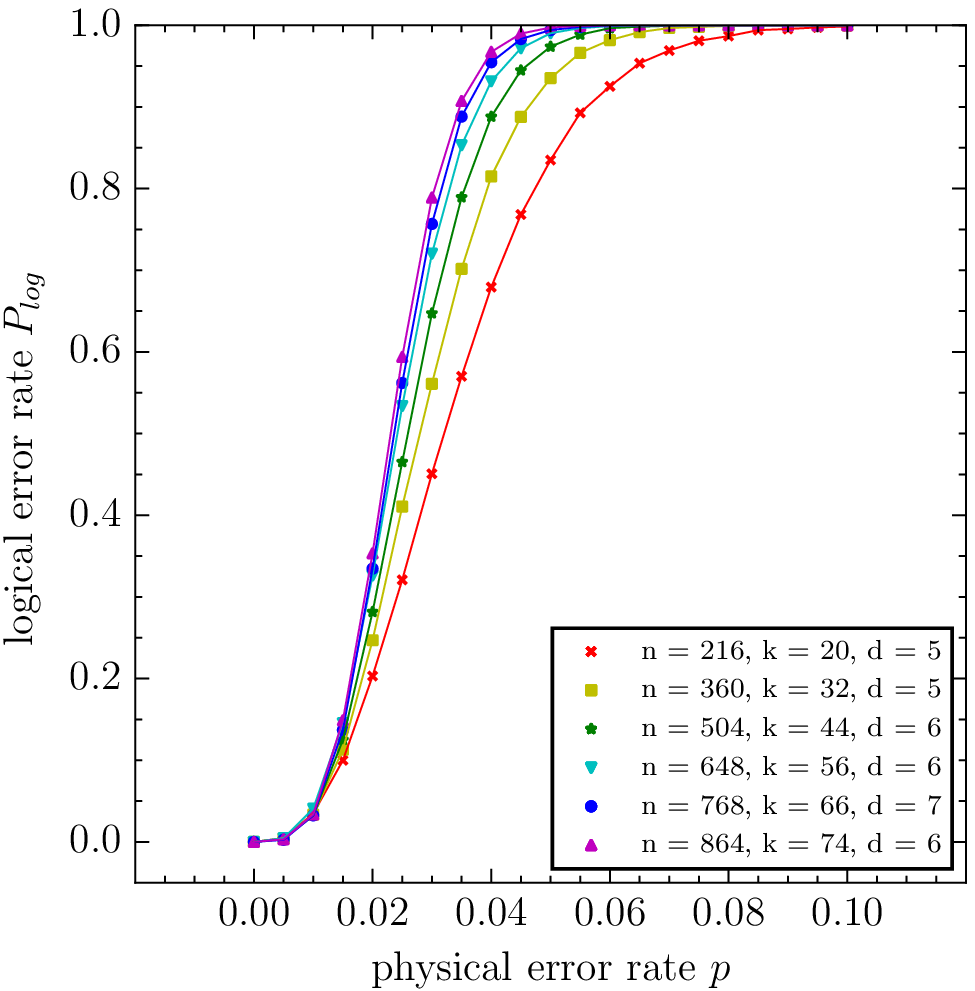}
      \caption{\schl{8}{3}-tiling}
      \label{fig:P_p_8_3}
    \end{subfigure}

    \begin{subfigure}{.43\textwidth}
      \centering
      \includegraphics[width=\linewidth]{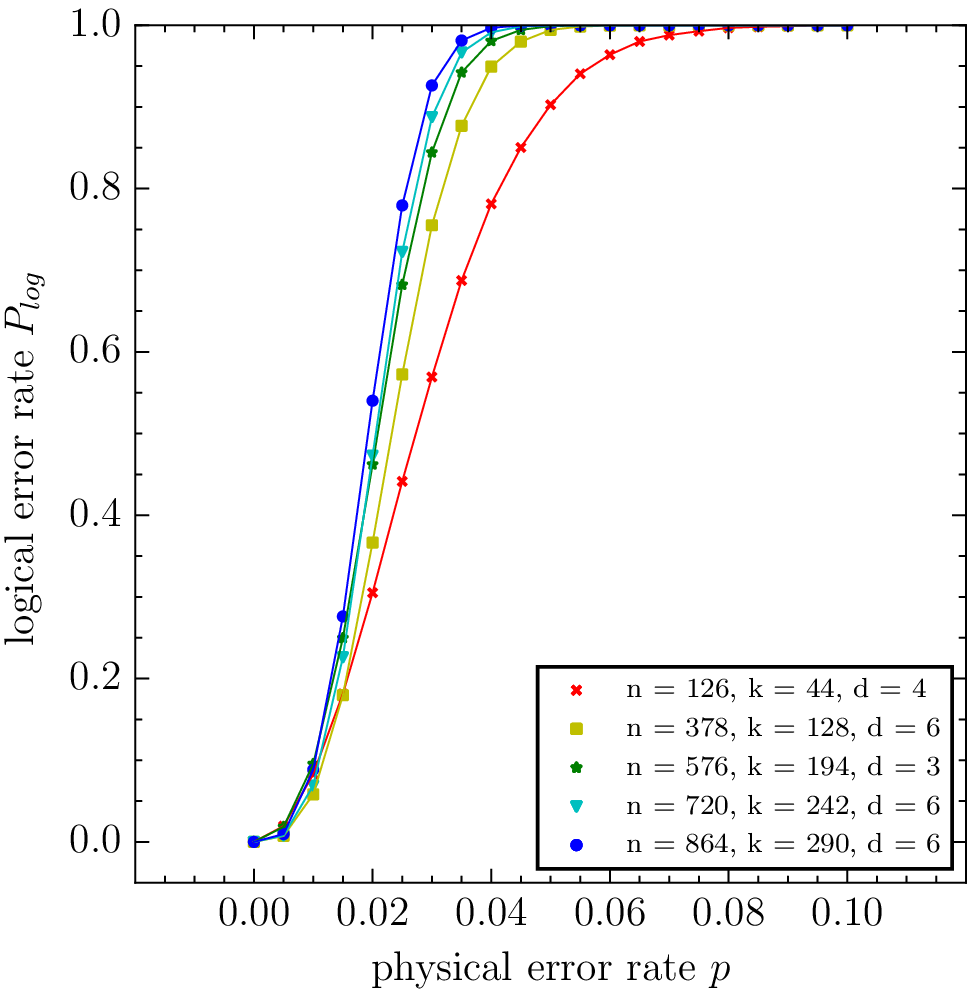}
      \caption{\schl{6}{6}-tiling}
      \label{fig:P_p_6_6}
    \end{subfigure}
  \caption{Logical error $P_{log}$ against physical error $p$ for hyperbolic surface codes with higher weight. Each data point was obtained by $N=4\times 10^4$ trials.}\label{fig:P_p_high}
\end{figure}

\begin{figure}[ht]
  \centering
    \begin{subfigure}{.43\textwidth}
      \centering
      \includegraphics[width=\linewidth]{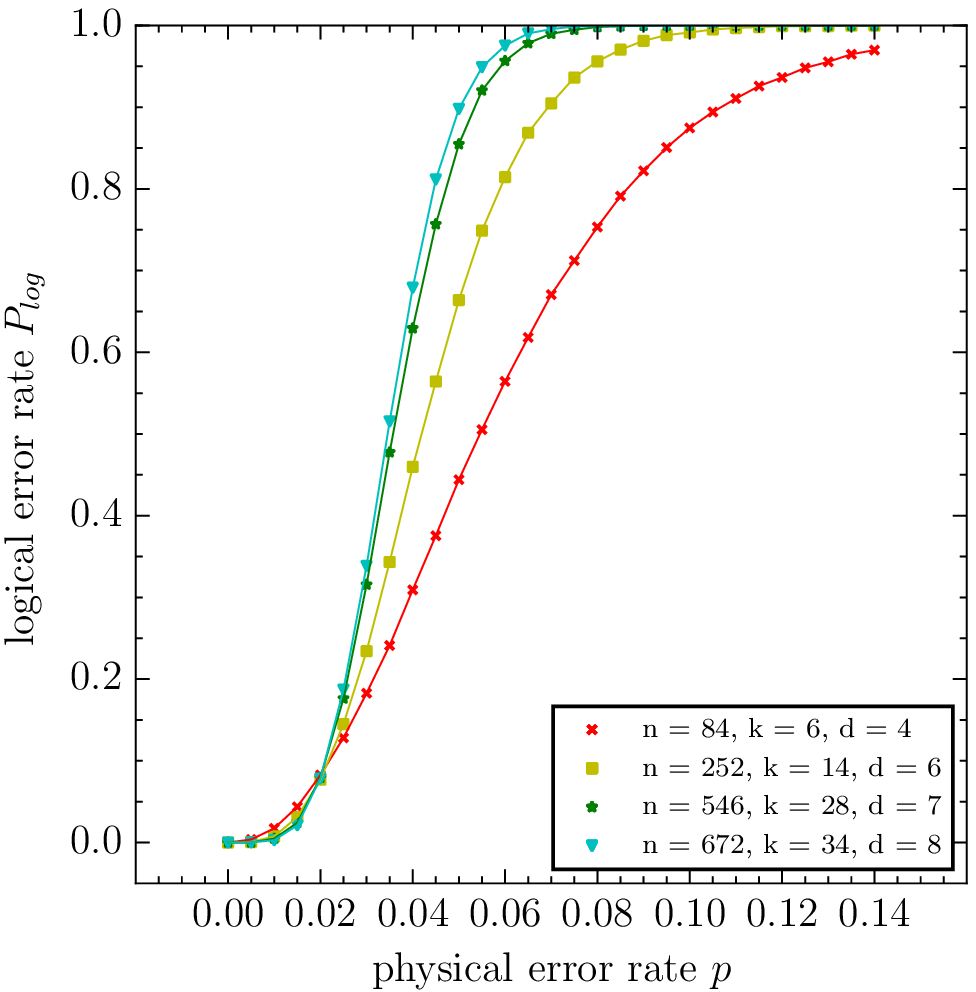}
      \caption{\schl{7}{3}-tiling}
      \label{fig:P_p_7_3}
    \end{subfigure}

    \begin{subfigure}{.43\textwidth}
      \centering
      \includegraphics[width=\linewidth]{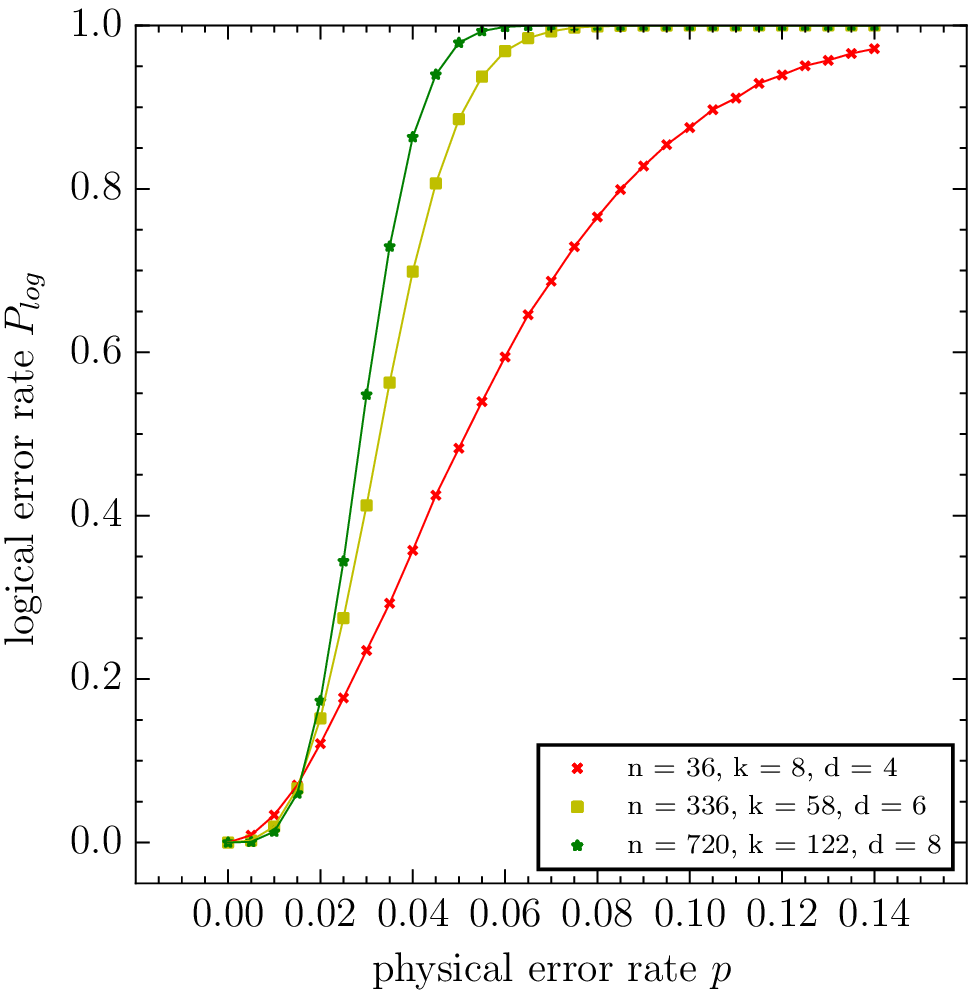}
      \caption{\schl{6}{4}-tiling}
      \label{fig:P_p_6_4}
    \end{subfigure}
  \caption{Logical error $P_{log}$ against physical error $p$ for hyperbolic surface codes with higher weight. We chose the examples to be strictly increasing in the distance. Each data point was obtained by $N=4\times 10^4$ trials.}\label{fig:P_p_high}
\end{figure}

\end{document}